\documentclass[aps,twocolumn]{revtex4-1}
\usepackage{amsmath,amssymb}
\usepackage{slashed}
\usepackage{graphicx}
\usepackage{bm}
\usepackage[T1]{fontenc}
\usepackage[utf8]{inputenc}
\usepackage{gauss} 
\usepackage[top=1.0in,bottom=1.0in,left=1.0in,right=1.0in]{geometry}
\usepackage[colorlinks,linkcolor=blue,citecolor=blue]{hyperref}
\usepackage{comment}
\usepackage{ytableau}
\newcommand{\be}{\begin{equation}}
\newcommand{\ee}{\end{equation}}
\newcommand{\bea}{\begin{eqnarray}}
\newcommand{\eea}{\end{eqnarray}}
\newcommand{\ba}{\begin{eqnarray}}
\newcommand{\ea}{\end{eqnarray}}

\begin{document}

\title{Pentaquarks made of light quarks and their admixture to
baryons
}

\author{Nicholas Miesch}
\email{nicholas.miesch@stonybrook.edu}
\affiliation{Center for Nuclear Theory, Department of Physics and Astronomy, Stony Brook University, Stony Brook, New York 11794--3800, USA}

\author{Edward Shuryak}
\email{edward.shuryak@stonybrook.edu}
\affiliation{Center for Nuclear Theory, Department of Physics and Astronomy, Stony Brook University, Stony Brook, New York 11794--3800, USA}

\author{Ismail Zahed}
\email{ismail.zahed@stonybrook.edu}
\affiliation{Center for Nuclear Theory, Department of Physics and Astronomy, Stony Brook University, Stony Brook, New York 11794--3800, USA}

\begin{abstract} 
This paper is a continuation of our studies of multiquark hadrons. The anti-symmetrization of their 
wavefunctions required by Fermi statistics is nontivial, as it mixes  orbital, color, spin and 
flavor structures.  In our previous papers we developed a method to find them based on the representations of the permutation group, and derived  the explicit wave functions for baryons excited to the first and second shells $(L=1,2)$, tetraquarks 
$qq\bar q\bar q$ and hexaquarks ($6q$). Now  we apply it
 to light pentaquarks ($qqqq\bar q$), in the S- and P-shells ($L=0,1$). Using Jacobi coordinates, one can use the hyperdistance approximation in
 12-dimensional space.
We further address the issue of ``unquenching" of baryons, by considering their mixing with pentaquarks, via two channels, through the addition of $\sigma$-like or $\pi$-like $\bar q q$ pairs.
This mixing is
central for understanding of the observed flavor asymmetry of the antiquark sea, the amount of orbital motion issue as well as  other nucleon properties. 
 \end{abstract}

\maketitle

\section{Introduction}
\subsection{Quark models of  hadronic structure: brief overview}
Well before the experimental and theoretical development of QCD, in the early 1960's Gell-Mann and Levy \cite{Gell-Mann:1960mvl} introduced the  concept of chiral symmetry. Its spontaneous breaking leads to  ``constituent quarks", as was shown by
Nambu-Jona-Lasinio (NJL)~\cite{Nambu:1961fr}. The lowest energy  mesonic states were three (nearly massless) pions and their chiral partner sigma, appearing in a chiral invariant potential depending solely on the invariant combination $(\vec \pi^2+\sigma^2)$. 

The discovery of QCD and heavy quarkonia in the 1970's, shifted the discussion 
to  gluon-exchange forces and confinement. Light quark spectroscopy was
dominated by ``bag models", e.g.  the  MIT bag \cite{Chodos:1974je},
the sigma bag~\cite{Friedberg:1977xf} and the ``little bag"~\cite{Brown:1979ui}, combining the MIT bag inside a pion-based Skyrmion. 

 Recent developments in hadronic spectroscopy is due to studies of multi-quark hadrons.
While discussed already in the 1960's, it has
received  considerable  experimental boost in the last decade. 
The clearest cases of multi-quark hadrons are  
charmed $tetraquarks$  $\bar c c\bar q q$. The first
 ``non-charmonium" $X(3872),J^P=1^-$ state, presumed to be a ``molecular" (deuteron-like) $DD^*$ two-meson state, just below its threshold. More tetraquark states have been reported later, e.g. the doubly-charmed
 $c c \bar q \bar q$ ones as well as   fully-charmed $\bar c \bar c c c$.
 %Since that time, many more tetraquark states of such composition were observed,  complemented by $c c \bar q \bar q$ ones, and even fully-charmed $\bar c \bar c c c$ ones.
 Originally called
$X,Y,Z$ etc for their unknown structures,  the current particle data group (PDG)  terminology splits $\bar c c\bar q q$ states into two categories.  Those with quantum numbers $I=1,J^P=1^+$
(pion-like $\bar q q$) are obviously tetraquarks,  now  appropriately called  $T_{\bar c c}$.
For  those with zero isospin,  vectors $J^P=1^-$ and pseudoscalars $J^P=0^-$ (with
 vacuum quantum number of $\bar q q=\sigma$ added), particle Data Group (PDG)  still uses their  generic notations, $\psi,\eta_c$ respectively, which  may be confusing. Yet at least four states  listed as $\psi$  in PDG 2024
  are tetraquarks, as they do not fit to be charmonia levels.

  Similarly, the
 addition of a $\bar q q$ with $\sigma,\pi$ quantum numbers to $qqq$ baryons would result in the pentaquarks. The history of their
 experimental and theoretical studies is full of puzzles we will not be able to cover it in full.
 
Isgur and collaborators~\cite{Kokoski:1985is,Geiger:1989yc} 
have argued that adding extra $\bar q q$ with 
 $\sigma$  (vacuum) quantum numbers require them to be in $^3P_0$
(J=0,S=1,L=1) state, following earlier suggestions such as \cite{LeYaouanc:1972vsx}.
This step stressed why certain orbital motion
in nucleons may be unavoidable.

Their pentaquarks were modeled via multiple baryon-meson states. Further studies along these lines were  carried out by \cite{Santopinto:2007aq,Bijker:2009up}) and others, resulting in the ``unquenched constituent quark model" (UCQM). 
 These works lead to realization that
all baryons, even the nucleons,  do contain 
a significant admixture ($\sim 40\%$) of 5-quark states.
(Original textbook results from the 3-quark constituent  models 
(e.g. magnetic moments) could still be preserved
due to certain cancellations in the 5-quark sector,
and new (e.g. orbital motion) can be explained.

Attempts to add a ``pion cloud" to the nucleon followed a different direction. An extreme case was
the ``Skyrmion" model, in which  the $qqq$ core is
abolished  and the baryons are entirely  made of the semiclassical pion fields. Less extreme approaches add pions via various
diagrams derived from effective chiral Lagrangians. 

We will treat the addition of $\sigma$-like and $\pi$-like
quark pairs on equal footing, as chiral symmetry suggests. Relations of 
 the admixture operators for them go back to 1950's, e.g. the Goldberger-Treiman relation and Gell-Mann-Levy ``linear sigma model"~\cite{Gell-Mann:1960mvl} as reviewed e.g. in~\cite{Nowak:1996aj}. 
%Yes, we need to admix $both$ $\bar q q$ combinations, and their 
%couplings  are indeed the same.

\subsection{Bridging hadronic spectroscopy and partonic observables on the light front}
The alternate direction to hadronic structure originated in the 1970's, and was  based   on data from Deep Inelastic Scattering (DIS) experiments, providing  Particle Distribution
Functions (PDFs). When expressed in terms of gluons and individual quark flavors,
PDFs are not wave functions but {\it density matrices}, traced over all variables except the longitudinal momentum of a single parton.  By definition, PDFs include
all multiquark and multigluon sectors. Their off-forward generalizations in the
form of Generalized Parton Distributions (GPDs) are actively sought both
theoretically and experimentally, and form the core program of the JLab facitility
and the future  Electron Ion Collider (EIC).

Data analysis further revealed their dependence on scale of momentum transfer
$Q^2$,  in agreement with the QCD predictions of anomalous dimensions
and/or ``perturbative evolution".  Unlike the chiral mechanism with the
addition of $\bar q q$ pairs mentioned above, they are
based on pQCD processes such as $g\rightarrow \bar q q$. While dominant
at large $Q^2\gg 1 \, GeV^2$, they do not generate the quark sea flavor asymmetry,
or quark orbital motion,  which are the main focus of the present paper.

\begin{figure}[h!]
    \centering
\includegraphics[width=0.95\linewidth]{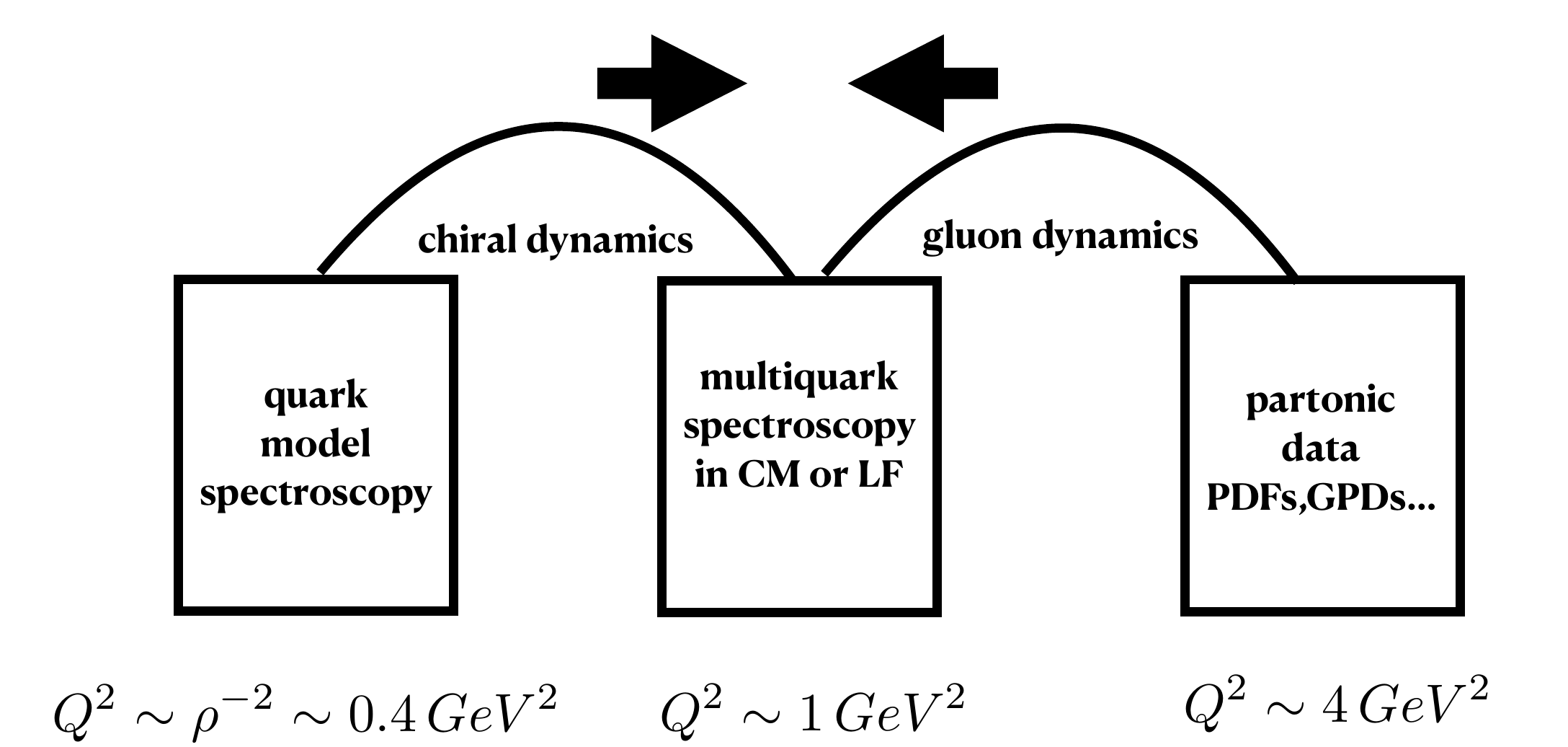}
    \caption{Bridging quark spectroscopy and partonic data}
    \label{fig_bridge}
\end{figure}

The main objective of our series of works is ``bridging" hadronic spectroscopy 
(as defined with Hamiltonians and wave functions,  either in the in CM frame or 
via light-cone formulation) to the  available parton data.  The main idea, in the form  of a ``double-ark bridge", is sketched in Fig.\ref{fig_bridge}.
 The intermediate description at the scale $Q^2\sim 1\, GeV^2$
should be derivable $both$ from the left via  ``chiral evolution"
(sigma-pion emissions), and  from the right, via pQCD evolution. 
 ``Bridging"  implies that these two different descriptions would eventually join smoothly at such intermediate
 scale. In particular, the mixed $qqq$ and $qqqq\bar q$  wave functions we will discuss in this work, would describe higher Fock components and the partonic observables,  explaining
puzzles associated with them.
%
% The ``intermediate pillar" in Fig.\ref{fig_bridge} is supposed to have only $one$ $\bar q q$
% pair admixed to a baryon. Unlike the (experimentally available) parton distributions  at high normalization scale,  at the matching pillar in the middle 
% the antiquark sea distribution should not have antiquark densities diverging at $x\rightarrow 0$, as those are due to gluonic processes $gluon\rightarrow \bar q q$.
% Its antiquark densities should be finite: but they still should keep the same flavor asymmetries as observed, as those are generated by chiral dynamics, from the left pillar.
% We also expect that the resolution of the spin-orbital puzzle will come from the left,
% generated by the fact that the vacuum $\bar q q$ pair include the orbital motion,
% as will be detailed below. 

%Of course, ``unquenching" the quark model by adding  $\bar q q$ pairs to $qqq$ baryons  has a rather long history.
%One  idea from 1970's was that a vacuum $q\bar q $ pair is
%in $^3P_0$  (S=1,L=1,J=0) state \cite{Micu,LeYaouanc:1972vsx}. 
%In 1990's it was included by Isgur and collaborators \cite{Geiger:1991qe}
%into a quark model. 
%Reviews on subsequent works adding a ``vacuum pair" and %considering
%intermediate states as $B+\pi$ in CM frame are e.g. in %\cite{unquenching}. 

%One nontrivial property of the sea (anti)quarks is their apparent {\em flavor asymmetry}, $\bar d(x)-\bar u(x)>0 $ ,
%in apparent disagreement with early applications of gluonic
%pQCD diagrams.

Adding ``pion clouds" to baryons via chiral diagrams  (such as $p\rightarrow n\pi^+$) 
generates a flavor asymmetry. It has a long history, for a recent review see \cite{Miller:2022nqi}.

Our works in this direction started with \cite{Shuryak:2019zhv}, with the addition of  a $\bar q q$  
to a nucleon  via instanton-induced 4-fermion t'Hooft Lagrangian. It was done in the light front (LF)
formulation, and  was able to reproduced not only the flavor asymmetry, but also the shape of the sea quark PFDs. Further discussion of the relative importance of the pion-based  and t'Hooft-based mechanisms, in  \cite{Shuryak:2022wtk}, concluded that their contributions are about equal. Hence both mechanism are needed to explain the data.

Another (highly publicized) issue is the ``spin puzzle". The naive
view that the nucleon spin is carried exclusively by the
valence quarks was disproved by experiments. 
The most important additional contribution is believed to be the $orbital$
motion in the nucleon. In our recent work 
\cite{Miesch:2025vas} we have investigated whether
it can be induced by a ``deuteron-like" mechanism, via
mixing with $D-shell$ baryons. We concluded that it is highly unlikely, 
even by assuming some extreme tensor forces.
In the current work, devoted to the admixture to pentaquark
states, the data about orbital motion seems to be 
reproduced successfully.

For completeness, we note that other authors continue to work on
 5-q Fock component of the nucleon as well, aiming to 
resolve  the same issues.  Perhaps the closest to our work is 
\cite{An:2019tld} in which  states of the $L=1$ pentaquark shell are discussed, 
including the mixing to the nucleon through the $^3P_0$ model.
%well as projecting a sigma-like perturbation to it. 
As the reader will see, our general construction is technically quite different, 
where the pentaquark wave functions are solely derived through the strictures of  Fermi statistics, rather than combinations of the representations of the color-spin-flavor groups. This latter construction is discussed in Appendices for comparison and completeness. More importantly,
and in contrast to ~\cite{An:2019tld},  our mixing of the nucleon and pentaquark states will enforce chiral symmetry.

What is different in this paper in comparison to the existing literature, is that
we will   %pushing the theory  of pentaquarks as far as technically possible. 
%nucleon wave function  are done differently. We do $not$ use an oscillator basis and notion of some baryon-meson
%pairs moving with momentum relative to each other. Instead  we consider  full set of 
 explicitly derive the wave functions of the
pentaquark states, at the S-shell ($L=0$) and the $P-shell$ ($L=1$). The key idea is
that one has to enforce the Fermi  statistics via representation of the permutation group $S_4$. These wave function are originally defined
 by tensors, with explicit orbital-color-spin-flavor indices. Then they are ``flattened", transformed  to vectors in multidimensional space of all possible ``monoms". The individual
coefficients are not numbers but 
functions of coordinates, which can be
naturally defined and manipulated 
by analytic tools provided by Wolfram Mathematica.

%So, one goal of this paper is to give 
The kinematics of  pentaquark
wave functions (first in the CM and eventually in light-front (LF) formulations) starts with
 4 Jacobi coordinates,  after elimination of the center-of-mass motion. We further work
with quarks of the same mass, all being light $u,d$, so the kinetic energy is
a Laplacian operator with a hypercentral symmetry. 

 While the pentaquark masses and WFs are of interest by themselves, here we use them mainly as experimental manifestation in {\em baryon-pentaquark mixing}.
In \cite{Shuryak:2022wtk} we already introduced the phenomenological part of this problem, especially the so called flavor asymmetry of the
proton/neutron antiquark sea. Yet the evaluation of its magnitude was done in ``DGLAP-like" approximation, in which the probability to get an extra $\bar q q$ pair
is calculated from the lowest order diagrams, ignoring any interactions 
or interferences between the produced and already existing quarks. 
Now we explore  the mixing of the 3-q and 5-q sectors in a theoretically more satisfactory way.

\section{Pentaquarks}
\subsection{Brief history }
The experimental searches for light pentaquarks were performed  for decades, but in so far have not resulted in well established  results.

Looking for a state with a flavor-distinct -- strange -- antiquark, the
 experiment \cite{LEPS:2003wug} announced  in 2003 a $\Theta=uudd\bar s$ resonance with mass and width $$M_\Theta=1.54\pm0.01\, GeV,\,\,\, \Gamma_\Theta < 25\, MeV$$
It was seen in few other experiments, but then was not found in  other
experiments with larger statistics.  For a recent review of the experimental situation and current plans see e.g.
\cite{Amaryan:2025muw}.
The chiral soliton model \cite{Diakonov:1997mm}
has predicted such light $\Theta$, as member of a anti-decuplet family. Multiple other
models were proposed, for recent discussion see e.g. \cite{Praszalowicz:2024mji}.

At that time, two of us had suggested a schematic quark-diquark
symmetry \cite{Shuryak:2003zi}, which suggested pentaquark structure approximated by a three-body object with  two ``good diquarks" $(ud)^2 \bar s$, rotating (in a $P$-wave).  In such a model it should be approximately degenerate with P-wave excited decuplet baryons, suggesting higher mass $M\sim 1.9 \, GeV$.

Needless to say, now that
we are far from such a naive model, with a full set of wave functions as derived below. While
 at this time there is no firm association of the derived states with identified resonances, we trust
 the general principles, Fermi statistics and the hyperdistance approximation. We simply follow the same path as in our previous studies of excited baryons, tetraquarks and hexaquarks.   
% Last but not least, 
These results are then  used as an intermediate ``stepping stones", leading to understanding of 5-q
Fock components of the nucleons.
% A qualitative conclusion of our and
% similar works is that the 3q-5q mixing
% is not small. Even for the nucleon ground state, the admixture amounts to
% $\sim 40\% $ probability. It should be
% even larger for excited baryons, but a lot of work needs to be done to
% elucidate those.

As usual,
 understanding of complicated  multiquark wave functions starts with
extreme quantum numbers, e.g.  the channel with maximal
 spin $J^P=5/2^+$. Before we discuss it theoretically, let us comment that perhaps it is  the resonance $N^*(2000)\,\, 5/2^+$,
seen in channels such as $N\pi,N \sigma, \Delta \pi,\Lambda K^*$, see the large multichannel analysis in~\cite{Anisovich:2011fc}.
It is  the second resonance with such quantum number, well above the first $5/2^+,N^*(1680)$ traditionally associated with the $qqq, L=2$ D-shell state. (Furthermore, it  is very close to the negative parity $N^*(2060)5/2^-$ state, perhaps its  chiral partner. )

 Finally, let us comment on pentaquark states with charm quarks. Our use of the hyperdistance approximation for fully-charmed tetraquarks \cite{Miesch:2024fhv}, 
 does reproduce the splittings between these three states. Unfortunately it cannot be used for $uudc\bar c$ pentaquarks  seen by LHCb, and  therefore will not be discussed here.

\subsection{Kinematics}
The coordinates of five bodies, with the CM motion excluded, are described by  four three-dimensional Jacobi coordinates,
% $\alpha,\beta,\gamma,\delta$ 
 \ba \label{eqn_penta_Jacobi}
 \vec X_1 &=& ( 15 \sqrt{2} \vec\alpha + 5 \sqrt{6} \vec\beta + 
    5 \sqrt{3} \vec\gamma + 3 \sqrt{5} \vec\delta)/30, \nonumber \\
 \vec X_2 &=&   
  (- 15 \sqrt{2} \vec\alpha + 5 \sqrt{6} \vec\beta + 
    5 \sqrt{3} \vec\gamma + 3 \sqrt{5} \vec\delta )/30,  \nonumber\\
 \vec X_3 &=&   ( - 10 \sqrt{6} \vec\beta + 5 \sqrt{3} \vec\gamma + 
    3 \sqrt{5} \vec\delta)/30,  \nonumber \\
 \vec X_4 &=&   ( - 5 \sqrt{3} \vec\gamma + \sqrt{5} \vec\delta)/10,  \nonumber\\
 \vec X_5 &=&   - (2 /\sqrt{5}) \vec\delta
\ea
with  the inverse relations as 
\ba \vec\alpha &=& ( \vec X_1 - \vec X_2)/\sqrt{2}, \nonumber \\
\vec\beta &=&
 ( \vec X_1 + \vec X_2 - 2 \vec X_3)/\sqrt{6} \nonumber \\
\vec \gamma &=& 
 \sqrt{3}/6 (\vec X_1 + \vec X_2 + \vec X_3 - 
    3 \vec X_4), \nonumber \\
   \vec \delta &=& 
 \sqrt{5}/2 ( \vec X_1 + \vec X_2 + \vec X_3 + \vec X_4)
\ea
The Jacobi coordinates are defined in a sequence, with the first two coinciding with 
the $\vec\rho, \vec\lambda$ traditionally used in baryon spectroscopy.

Each 3d vector can in turn be parameterized by its magnitude and two polar angles $\theta_i,\phi_i,i=1,2,3,4$. The 4d space of
their magnitudes $|\alpha|,|\beta|,|\gamma|,|\delta|$
are described by the $hyperdistance$ in 12d
$$ Y^2=
\vec\alpha^2+\vec\beta^2+\vec\gamma^2+\vec\delta^2$$
and three new angles called $\theta 1_\chi, \theta 2_\chi, \phi_\chi$.
In total there are  $Y$, 8 ``ordinary" angles and 3 ``extraordinary" ones. 
The integration measure can be defined by a product of four 3d and one 4d solid angles 
\be d\Omega_{11}= d\Omega_\alpha d\Omega_\beta d\Omega_\gamma d\Omega_\delta d\Omega_\chi \ee
with $d\Omega_\chi=\rm 
sin^2(\theta2_\chi) sin(\theta1_\chi) d\theta2_\chi d\theta1_\chi d \phi_\chi $ and
 the usual $d\Omega_i=\rm\sin(\theta_i)d\theta_id\phi_i$.
 The generic wave functions can be derived 
in a factorized form 
$$ \Psi= R_n(Y) F(\theta1_\chi,\theta2_\chi,\phi_\chi) \Pi_{i=1}^4 Y_{l_i,m_i}(\theta_i,\phi_i)$$
Further details on the induced geometry -- the 
metrics and Laplacian -- are given in Appendix~\ref{sec_5d}.

\subsection{The radial wave functions in hyperdistance approximation }
%\section{Pentaquarks  in the CM frame}

% The total dimension of space is thus 12. One parametrization of it uses magnitude and directions of Jacobi vectors
% as $$\vec \alpha=\alpha \cdot \hat \vec n_\alpha(\theta_i,\phi_i), ...$$
%with modulus and unit vectors $\hat n_i^2=1$  with two angles each. 
%The four moduli  $\alpha,\beta,\gamma,\delta$ can be alternatively described by spherical coordinates,
%with hyperdistance \be Y^2=\alpha^2+\beta^2+\gamma^2+\delta^2 \ee and three angles one may
%call $\chi,\xi,\omega$. Thus the solid angle can be defined as  
%\be d\Omega_{11}\sim d\Omega_\alpha d\Omega_\beta d\Omega_\gamma d\Omega_\delta sin^2(\chi) sin(\xi) d\chi d\xi d \omega \ee

We  make the standard simplifying assumption that the ground state of pentaquarks made
of the same-mass quarks,  is approximately spherically symmetric.
In Jacobi coordinates the kinetic energy is proportional to the Laplacian
described in full in Appendix~\ref{sec_5d}.
For a spherically symmetric function depending only on the hyperdistance, it is
\be K=-{\Delta \over 2m}=\big(-{\partial^2 \over \partial Y^2}- 
{11 \over Y }{\partial \over \partial Y} + \Delta_\Omega \big){1 \over 2m}\ee 
The standard re-definition of the radial wave function,
and Laplacian eigenvalues for $L=0,1$ states, are
 given in Appendix~\ref{sec_5d}.

\begin{figure}[h]
    \centering
    \includegraphics[width=0.8
\linewidth]{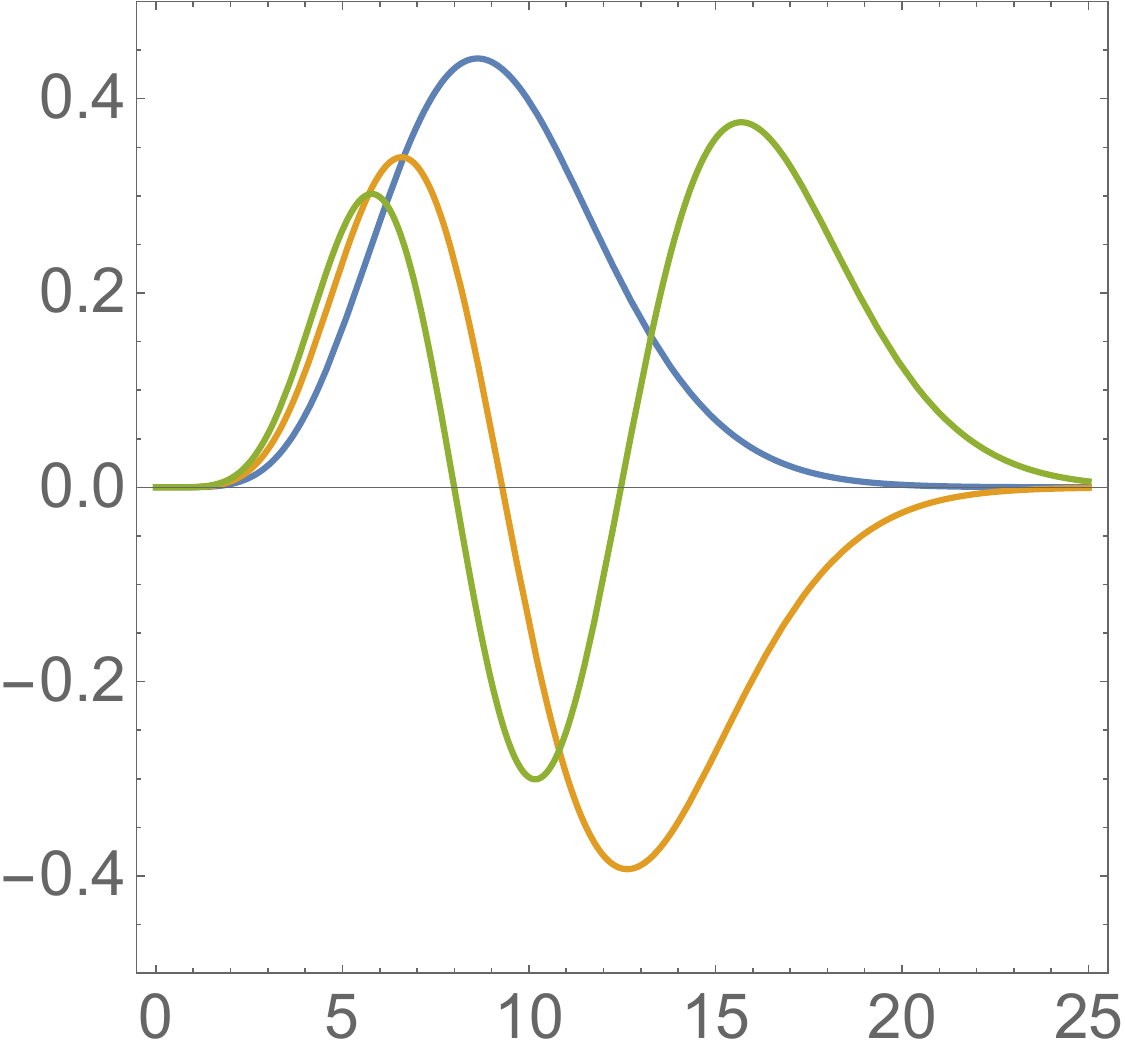}
    \caption{We show the two lowest spherical wave functions from  the radial Schrodinger equation, normalized as $\psi(Y)*Y^{11/2}$, versus the hyperdistance $Y$ in $\rm GeV^{-1}$. The solid lines are for the spherical $L=0$ states, and the dashed lines are for $L=1$ states.}
    \label{fig_penta_WFs}
\end{figure}

The potential energy is usually
a sum of binary potentials, which are $not$ spherically symmetric.
For instance,  the Cornell-like potential is projected to $\alpha,\beta,\gamma,\delta$ (all positive) fraction of a sphere, see Appendix~\ref{sec_5d}. Using it to solve the radial equation, we obtain the wave functions shown in Fig.\ref{fig_penta_WFs}. Note that for the ground state, the wave function  peaks at $Y\approx 10\,\rm GeV^{-1}\approx 2 \, fm $. (Recall that $Y^2$ is $not$ the squared interquark distance, but 1/5 of sum of squared distances between all
10 pairs.)

The calculation of  the radial hyperdistance wave functions for baryons
and tetraquarks,  was done in several papers 
(including ours \cite{Miesch:2024vjk,Miesch:2025vas}). The same procedure 
will be used  for pentaquarks. 
The light quark mass was taken as $M=0.35\,\rm  GeV$, and the potential is Cornell-type with proper averaging over angles. In order to avoid the unknown
constant shift in the potential, we only
use as predictions ``splittings" (differences)
between energies, e.g. for the
five lowest $L=0$ pentaquark masses, we have
\be \label{eqn_gaps}
E_n(L=0)-E_0(L=0)= 0.32, 0.62, 0.89, 1.16 \,\, \rm GeV \ee

 The $L=1$ radial equation includes the extra centrifugal
 term in the Laplacian $10/Y^2$, the corresponding splitting is
 $E_2(L=1)-E_1(L=1)= 0.30 \,\, GeV$.  The corresponding wave functions are shown in Fig.\ref{fig_penta_WFs}.

\subsection{  Pentaquarks with maximal spin $S=5/2$}
Before we discuss the case at hand, the pentaquarks $q^4\bar q$, 
let us remind  the logic used in \cite{Miesch:2024vjk} for
a related case of 
hexaquarks  ($q^6$).
 As usual, one starts with 
the simplest cases, e.g. those of maximal possible spin, $S=3$ for hexaquarks or $S=5/2$ for pentaquarks.
Recall that the $S=3$
   hexaquark is 
 the only case for which
 a fully anti-symmetric WF has actually been derived analytically
 in~\cite{Kim:2020rwn}. It was done  by traditional means,
 adding up available color and flavor representations. The answer was written explicitly,
as a sum of  five parts with different color-flavor structures. 
This state is believed to be
  experimentally observed~\cite{WASA-at-COSY:2011bjg} as a resonance $d^*(2380)$ in reaction
$$ p+n \rightarrow d +\pi^0 +\pi^0$$ 
 Its  width $\Gamma_{d^*}\approx 70\, MeV$ is significantly below that of Delta baryon $\Gamma_\Delta\approx 115\, MeV$, which   was used against its interpretation as a $\Delta \Delta$ bound state. Also,  for a 
 ``deuteron-like" interpretation of a $\Delta \Delta$ state, the binding needs  to be $\approx 84\, \rm MeV$, which is perhaps too large.

Similarly, the simplest pentaquarks must be those with maximal possible spin $5/2$.  
If the isospin is also maximal,  the only function remaining is that of color, which cannot be antisymmetric since the color Young tableaux height is restricted to $N_c=3$. Yet the antisymmetric pentaquark states with $S=5/2,I=3/2$
and $S=5/2,I=1/2$  can be constructed,
even
by the usual procedure of combining color and flavor
representations. Since its discussion is rather long, it is 
delegated to Appendix~\ref{app_analytic}. 

The wave function for $S=5/2,I=1/2$ state is given in (\ref{PENTA1}). We
translated the monom basis using Wolfram Mathematica, 
obtaining a 144-component wave function.
Compared to the one from alternative derivation based on 
permutation group (to be explained next) we
have  observed full agreement.

Some  properties of the maximal spin $S=5/2$ states
such as
\be \langle \sum_{i>j}\vec\lambda_i\vec \lambda_j \rangle=-40/3 \ee 
\be  \langle \sum_{i>j}(\vec\lambda_i\vec \lambda_j)(\vec S_i \vec S_J) \rangle   =-10/3 \ee
 obtained analytically, hold for other (much more complex) pentaquark states to be discussed below. Such  numerical checks
 turned out to be very useful.

The  distributions of flavor and color between the quark ``core" and the antiquark are nontrivial.
For instance,  for the $S=5/2,I=1/2$ state
the mean values of the isospin projections are \be \langle I_z(q)  \rangle=1/4,\,\,\, 
 \langle I_z(\bar q)\rangle =-1/2 \ee
(Note that the sum rule $4\langle I_z(q)  \rangle+\langle I_z(\bar q)\rangle=1/2$ holds as expected.) 

%for $S=5/2,I=1/2$ the isospin averages are
%\be \langle I_z(q)  \rangle=1/6,\,\,\, 
% \langle I_z(\bar q)\rangle =-1/6 \ee

\subsection{ Color-spin-flavor wave functions  for  S-shell ($L=0$) pentaquarks from permutation symmetry}
  Signifiant technical difficulty in building the theory of multiquark
  hadrons is a need to fulfill permutation antisymmetry of all quarks required by
  Fermi statistics. The traditional way of doing it is by
  adding a pair of quarks into all possible representations 
  of  all pertinent groups ($SU(3)$ for color, $SU(2)$ for spin and flavor, etc), and then adding next quark (or a pair), etc.  How to do it is described  in Appendix~\ref{app_analytic}.
  
%%%%%%%%%%%%%%%%%%%%%%%%%%

\begin{table}[b!]
    \centering
    \begin{tabular}{|c|ccc|}   \hline
         I \textbackslash \, S &0 & 1 & 2   \\
         \hline
         0&0& 1& 0 \\
         1& 1&1 & 1 \\
         2&0&1&0 \\ \hline
    \end{tabular}
    \caption{Number of states (per choice of $I_z$ and $S_z$) for each combination of total spin and isospin for four quarks.  Adding an antiquark can shift each state's spin  up or down by 1/2.}
    \label{tab:4q}
\end{table}
%%    \centering
%    \begin{tabular}{|c|ccc|}   \hline
%         I \textbackslash \, S &1/2 & 3/2 & 5/2   \\
%         \hline
%         1/2&3& 3& 1 \\
%         3/2& 3&3 & 1 \\
%         5/2&1&1&0 \\ \hline
%    \end{tabular}
%    \caption{Number of states (per choice of $I_z$ and $S_z$) at each %combination of total spin and isospin after multiplying on the antiquark.--}
%    \label{tab:5q}
%\end{table}

The wavefunctions to be derived here follow from the novel technique we developed in \cite{Miesch:2024vjk}. For pentaquarks it is based
on finding representations of the 
``quark core" using the permutation group $S_4$. %Antisymmetric wavefunctions, permissible for
%Fermi statistics of 4q, can all be found explicitly. While we will derive it 
%below,
Let us start with the table \ref{tab:4q} giving the number of states of the ``core", as a function 
of isospin  $I$ and  spin $S$. (As expected, the table is symmetric under their interchange.) 

After the antiquark is added through standard procedure,  one gets the number of pentaquark states given in the  table \ref{tab_L0}. 
%Because it is not required to have exchange symmetry with the quarks and can be either up or down in spin or isospin, adding the antiquark allows for significantly more states.
The antisymmetric representations of the permutation
group for four light quarks 
 $S_4$, follows the procedure developed in
\cite{Miesch:2024vjk}. 
Standard ``monom space" for S-shell ($L=0$) pentaquarks has color-spin-flavor dimension
\be \label{eqn_monoms}
d_{monoms}=3^6 \times 2^5 \times 2^5=746496 \ee
(note that the antiquark in color Young tableaux is represented by two squares, thus $3^6$.) 
In the case of P-shell ($ L=1$) states (we will need for baryon mixing study below) it is additionally 
multiplied by 4, the number of Jacobi coordinates which appear linearly, the monom space dimension  extends well over a million. 
It may appear that, even with
the help of Mathematica, writing operators as matrices in this huge space would be  impossible.

Fortunately, this is not so, for two reasons. First, for each of 4 subspaces (orbital, color, flavor, spin) 
one can define and operate in what we call the ``good basis" based on pertinent Young tableaux.  Therefore in 
practice one can define the permutation 
generators as matrices in spaces of  dimensions much smaller than  $d_{monom}$
$$N_{GB}=N_{GB}^{color}\times N_{GB}^{spin}\times N_{GB}^{flavor}$$
  Two basic generators of the permutation groups are defined by the {\em ``KronekerProduct"} operator in Mathematica of respective sub-matrices, and diagonalized. The wave functions we look for given by 
 $antisymmetric$ eigenvectors {\em common to both generators}. Since
all $n!$ elements of $S_n$ group can be constructed out of those two generators,
such  wave functions are  antisymmetric  for all permutation of the group.

For spherically symmetric S-shell ($L=0$) states 
there are color*spin*flavor subspaces: the
antisymmetric states we  found are 
listed in Table \ref{tab_L0}. The number in brackets are dimensions of respective ``good basis" in each case.
%for example for $I=J=3/2$ it is $27=3\times 3 \times 3$.
While resulting permutation
matrices are still too large to be presented here, their dimensions are much smaller than that of the ``monom space 
(\ref{eqn_monoms}), and   dealing with them
 creates no problem.

Consider as an example the pentaquarks with
the maximal spin $S=5/2$.
The remaining nontrivial Young tableaux  are those of color and isospin.  Using the familiar
$SU(2)$ addition of five spins one readily finds that 
there are 5 ways to add them to total isospin $I=1/2$:
thus the ``good basis" for isospin has dimension 5. 
The
Young tableaux for color adding to total color zero
is made of two closed columns $\epsilon_{abc}\epsilon_{def}$, like for hexaquarks.
Unlike that case, however, permutations should be 
done for quarks only, 4 indices $abcd$. This leads to a
``good basis" of dimension 3, or $3\times 5=15$ in total. The matrices of two generators of $S_4$  are 
then defined and diagonalized. One $common$ eigenvector with eigenvalue $-1$ (antisymmetric) is found, out of 15. 
 therefore
we have found the $single$ pentaquark state with $S=5/2,I=1/2$. (Recall that for hexaquarks with maximal spin it also was a single state.) 

\begin{table}[h!]
    \centering
    \begin{tabular}{|c|c|c|c|} \hline
  I/S    & 1/2 & 3/2   & 5/2  \\  \hline
  1/2 & 3 (75) & 3 (60)   & 1 (15) \\
  3/2 & 3(60) & 3 (48) &  1 (12)  \\
  5/2 & 1 (15)  & 1 (12) & 0 \\ \hline
    \end{tabular}
    \caption{Spin $S$- isospin $I$ table
    of 
    antisymmetric pentaquark states
    with $L=0$. 
    The integer numbers  show the multiplicity of the states, the number in brackets are the
    dimensions of the ``good basis" used 
    to do calculations in each of the $S,I$ sectors.
   .}
    \label{tab_L0}
\end{table}

After all relevant  antisymmetric states are found,  the most efficient way to proceed is to
return to the universal description in the ``monom basis" of size (\ref{eqn_monoms}) for
all WFs.  While vectors and operators (matrices of this size squared, $d_{monoms}\times d_{monoms}$) may appear
frighteningly large, {\em Wolfram
 Mathematica} let us deal with them, with  the usual notations
e.g. $vector^*.matrix.vector$ \footnote{Note that while $SparseArray$ format for large vectors and matrices may speed things up,
in this form some general commands like Expand, Conjugate, ComplexExpand etc do not work (in version 14.1), thus it is safer to use $Natural$ format for such manipulations.
}.

The actual number of nonzero components of the WFs is in fact much smaller than
$d_{monoms}$,  so these arrays are really  quite sparse. Although we still cannot publish them
in a paper, and  only present average values of pertinent operators, it is still instructive
to compare the number of nonzero terms in the monom
representation. As anticipated, the smallest  
numbers are for maximal spin, $S=5/2$. The one with $I=3/2$ has only 252 nonzero terms, while that with
$I=1/2$ has 708 non-zero terms. Going to non-maximal spin 
and then to $L=1$ P-shell  leads to a significant increase in the number of nonzero components, e.g. 13 $L=1,S=1/2,I=1/2$ states we will use below (for mixing
with the nucleons) have 187200 nonzero components (together).

For $I=1/2$ and $S=1/2,3/2$ we found two triplets  states each, as can also be derived based on
Young tableaux (see Appendix~\ref{app_analytic}). 
The triplet states one gets are quite arbitrary, and  only the eigenstates of certain physical operators 
make them physically distinct. The sum over ten pair operators $(\lambda\lambda)=\sum_{i>j}\vec \lambda_i \vec\lambda_j$ (for definition of antiquark
color matrix see Appendix~\ref{app_analytic})
and $SS=\sum_{i>j}\vec \sigma_i \vec\sigma_j$
are related to the Casimirs of color and spin, so they are
diagonal for these states. One needs a Hamiltonian, e.g. with
 one-gluon exchange color-spin
forces 
summed over
 all 10 pairs
\be   \label{eqn_H}
H_{\lambda\sigma}=-C_{\lambda\sigma}\sum_{i>j} (\vec\lambda_i \vec\lambda_j)(\vec S_i \vec S_j)\ee
We evaluaed the full Hamitonian (matrix elements between all states)
and diagonalized it. The obtained
eigenvalues  for seven of them for $I=1/2$ ($S=1/2$ triplet, $S=3/2$ 
triplet, and $S=5/2$ singlet) are given in Table~\ref{tab_L0_properties}.
As expected, the state with maximal spin $S=5/2$
has also the largest energy.
(If
$N^*(2000) 5/2^+$ happen to be indeed this pentaquark,
others are then expected to be somewhat below $2\,\rm GeV$ in mass.)

(The alternative Hamiltonian, used e.g. by \cite{An:2019tld}, is the flavor-spin
operator corresponding to pion exchanges, following Ripka and Glozman
\be    H_{\tau\sigma}=-C_{\tau\sigma}\sum_{i>j} (\vec\tau_i \vec\tau_j)(\vec S_i \vec S_j)\ee
Note that for spin we use $\vec S=\vec\sigma/2$
with a half, but for color and flavor we use
Gell-Mann and Pauli matrices $without$
a half. Having the state wave functions,
we also calculated the Hamiltonian matrix for it. We do not present its eigenspectrum
as it is not qualitatively different from
that of color-spin Hamiltonian.)

Due to symmetry, the mean value of the $S_z$ component is the same for all four quarks, but different for the antiquark. 
By construction, the mean values of $S_z(i)$  are related by the sum rule 
\be \label{eqn_S_sumrule}
\sum_{i=1..5}\langle S_z(i) \rangle = 4\langle S_z^{1}\rangle+\langle S_z^{5}\rangle  = S_z^{total} \ee

% from penta0_S12.nb,penta0_S32.nb and 52_Ismail_mod.nb
%\begin{widetext}
\begin{table}[h!]
    \centering
    \begin{tabular}{|c|c|c|c|c|c|c|c|} \hline
  S & 1/2  & 1/2 & 1/2 & 3/2 & 3/2 & 3/2  & 5/2 \\  \hline
$\lambda \lambda$ & -40/3  & -40/3 & -40/3 &  -40/3  &  -40/3  &  -40/3  & -40/3  \\
$SS$ & -3/2  & -3/2 & -3/2 & 0 & 0 & 0 & 5/2  \\
 $H_{\lambda \sigma}/C_{\lambda \sigma}$ & 
-4.66 & -1.44 & 2.77 & -3 & 1/3 & 10/3 & 10/3 \\
% $\langle S_z^{q}\rangle$ & 0.167 & 0.097& 0.069 & 0.389 & 0.316 &  0.001 & 1/2\\
%  $\langle S_z^{\bar q}\rangle$ & -0.167 & 0.111 & 0.221 & -0.02 & 0.378 & 0.49 & 1/2 \\
%   $\langle I_z^{q}\rangle$ & 0. & 0.167 & 0.167 & 0. & 0.159  & 0. & 1/4\\
%  $\langle I_z^{\bar q}\rangle$ & 0.5 & -0.167 & -0.167& -0.170 & -0.153 & 0.5& -1/2  \\
 \hline
    \end{tabular}
    \caption{We list the properties of the triplet of $S=1/2$, the triplet of $S=1/2$ and single  $S=5/2$ states, all seven with the isospin $I=1/2$ (the upper line in the previous table).
   .}
    \label{tab_L0_properties}
\end{table}
%   \end{widetext}

The mean values of spins and isospins of the
quarks and antiquark can be compared to the expectation values based on naive models. In particular, if the pentaquark is made of two spin and isospin zero diquarks \cite{Shuryak:2003zi}, the $\langle S_z^{q}\rangle\approx \langle I_z^{q}\rangle \approx 0$ and all the spin and isospin
are carried by the antiquark $\langle S_z^{\bar q}\rangle \approx \langle I_z^{\bar q}\rangle \approx 1/2$. Compared to results in the table one can see some states do have either zero mean spin or zero mean isospin on quarks, but both never occurs. It is one more example showing how naive such models really are. 

The alternative naive model, in which the spin and isospin are divided equally between all 5
objects, also does not seem to be supported by this detailed analysis.

\subsection{ Pentaquarks of the $L=1$ shell}
Because the $\sigma,\pi$ operators introducing pertinent mesons (to be used below
to ``unquench" the nucleons) necessarily include nonzero orbital momentum,  the P-shell pentaquark states can mix with  the nucleon.

For $L=1$ pentaquarks, 
the dimension of ``good states" increases by a factor 4 
on top of what it was in S-shell, being just the number of Jacobi coordinate vectors $\vec\alpha,\vec\beta,\vec\gamma,\vec\delta$ appearing linearly. 
There are approximately that many more fully antisymmetric states found, see  
the list of the $L=1$ states in Table~\ref{tab_L1}. 

The states that can mix with the
nucleon should have $I=1/2$, and either $S=1/2$ or $S=3/2$. The list of Fermi-statistics-compliant states 
is given in the Table~\ref{tab_L1}, 24 in total~\footnote{Note that in~\cite{An:2019tld} the number of $L=1$ compatible states is 17, and they are not organized into Fermi-statistics-compliant combinations. We also disagree with the way their energies
are derived from their Hamiltonian.}. 

Note that even the largest ``good basis" with dimension 300, is still many orders of magnitude smaller than the ``monom basis" of dimension 746496.  
%Mathematica has no problem operating with permutation matrices and their eigenstates. Furthermore, for standartization of states and especially operators, we
%return to the full monom basis. 
The vectors are sparse but not too sparse: the 13 $S=12$ states
have 187200 nonzero matrix elements, 11 $S=3/2$ states 
have 198000 nonzero matrix elements. P-shell states are much
more complicated than the wave functions of the $L=0$ shell, as the amplitude
 is no longer a number but an explicit function
of 12 coordinates, e.g. eight angles $\theta_i,\phi_i, i=1,2,3,4$ and four magnitudes.  
The operators of angular momenta contain differentiations over angles, as usual,
which can be directly applied to our sparse vectors of very large dimension.

\begin{table}[h!]
    \centering
    \begin{tabular}{|c|c|c|c|} \hline
  I/S    & 1/2 & 3/2   & 5/2  \\  \hline
  1/2 & 13 (300) & 11 (240)   & 3 (60) \\
  3/2 & 11 (240) & 10 (192) &  3 (48)  \\
  5/2 & 3 (60)  & 3 (48) & 1 (12) \\ \hline
    \end{tabular}
    \caption{We list the antisymmetric pentaquark states
    with $L=1$, spin $S$ and isospin $I$. As in the previous Table \ref{tab_L0}, 
    the first integers  are the numbers of corresponding states, the numbers in brackets show the
    dimension of the corresponding ``good basis". }
    \label{tab_L1}
\end{table}

Some operators, like color $$\sum_{i>j}\vec\lambda_i \vec\lambda_j\rightarrow -{40 \over 3}$$ 
%and  \sum_{i>j}\vec S_i \vec S_j $
have the same expectation value for all states, owing to the $[211]$ color representation
of the Young tableau for the core.

\begin{figure}[h!]
    \centering
    \includegraphics[width=0.95\linewidth]{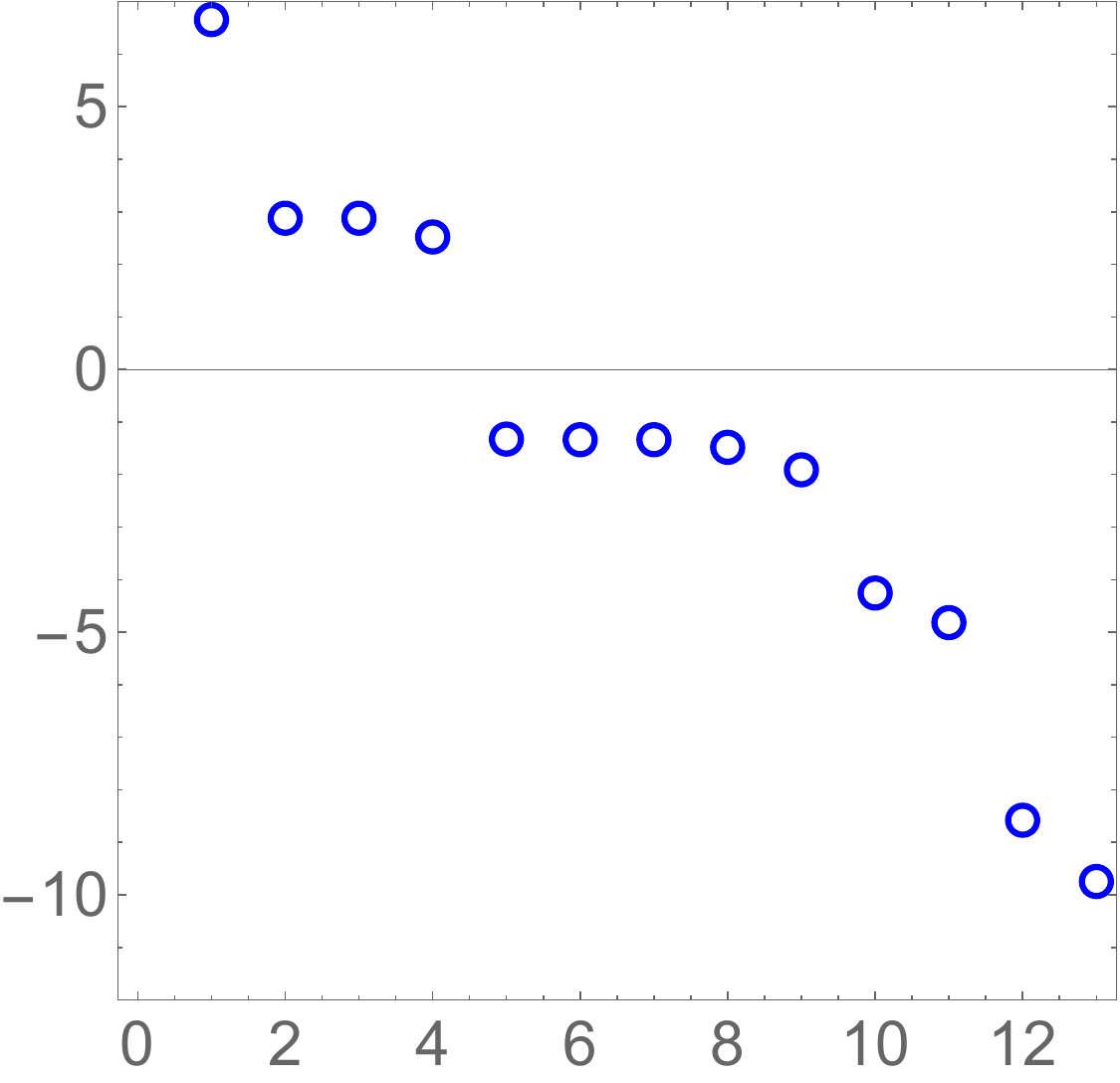}
    \caption{Eigenvalues of the color-spin Hamiltonian for $L=1,S=1/2$ states,  normalized to the matrix element of the potential $H_{\lambda\sigma}/C_{\lambda\sigma}$.
    }
    \label{fig_pentas_L1}
\end{figure}

The eigenvalues of the $13\times 13$ color-spin Hamiltonians for the $L=1$ pentaquark shell with
     $S=1/2$ are shown in Fig.\ref{fig_pentas_L1}.
%     While spectra turned out to be rather similar,
%     the states themselves may not be so similar. 
%  While we do not think it is likely that
%  those states can be experimentally individually identified due to overlapping widths, we will show below that certain combinations of them do constitute the 5-q Fock
%  sector of the nucleons.

% from pentas_operator_6.nb

As we did for $L=0$ pentaquark shell, we calculated
the distribution of various physical quantities over the 4 Jacobi solid
angles and the 5 bodies at hand. In order to not clutter the presentation,
we only present those for the three select states, numbers 1 ,7, 13 of the $S=1/2$ set in Table \ref{tab_3states}. 

The four Jacobi coordinates have four pairs of angles $\theta_i,\phi_i$, and so we calculated the expectation values
of  the orbital momenta $squared$, in each of the four Jacobi solid angles 
\be L^2_if= \rm -{1 \over sin[\theta_i]} {\partial \over \partial\theta_i} [sin[\theta_i] {\partial f\over \partial\theta_i}]  - {1 \over sin[\theta_i]^2} {\partial^2 f\over \partial\phi_i^2} 
  \ee
 $i=1,2,3,4$. The symmetry between the quarks in the core leads their equality $\langle L^2(1)\rangle =\langle L^2(2)\rangle=\langle L^2(3)\rangle$, but the corresponding  value for the antiquark --related to the fourth angle -- is different. 
As seen from this table, the three selected states  show a large
variety of distributions. The first state in the second column is the one of the lowest energy.
It has nearly no angular momentum on the antiquark, and nearly no spin on the quarks.

We also calculated mean values of $z$-component of $L_z(i),i=1..4$
on sum states.
All of our results reproduce  accurately the
respective sum rules
\be \sum_{i=1..4}\langle L^2(i) \rangle =3*L^2(1)+L^2(4)=2 \ee
\be \sum_{i=1..4}\langle L_z(i) \rangle =3*L_z(1)+L_z(4)=-1 \ee
as expected for the $L=1$ shell. 
(For completeness, we recall that the spin and isospin components  satisfy different sum rules 
$$ \sum_{i=1..5}\langle S_z(i) \rangle=4\times S_z(1)+S_z(5)=1/2 $$
$$ \sum_{i=1..5}\langle I_z(i) \rangle =4\times T_z(1)+T_z(5)=1/2 $$

\begin{table}[h!]
    \centering
    \begin{tabular}{|c|c|c|c|} \hline
  state number    & 1 & 7   & 13  \\  \hline
$L^2(1) $ & 0.4965 & 0.6652 & 0.6547 \\
$L^2(4)$ & 0.0258 & 0.0043 & 0.0358\\
$S_z(1)$ & 0.0963 & 0.0562  & 0.0931 \\
$ S_z(5)$& 0.1146 & 0.2748 & 0.1272  \\
$ I_z(1)$& 0.1523 & 0.0516 & 0.0545 \\
$I_z(5)$& -0.1094 & 0.2934 &  0.2817 \\ \hline
    \end{tabular}
    \caption{Distribution of few physical quantities over Fermi-antisymmetric pentaquark states
    with $L=1$, spin $S=1/2$ and isospin $I=1/2$. The operator $L^2(i)$ is the squared angular momentum,
    in the first and fourth Jacobi solid angles. The
    mean z-components of spin and isospin are given for the first quark and the 5-th antiquark, respectively.
    }
    \label{tab_3states}
\end{table}

\begin{figure}[t]
    \centering
        \includegraphics[width=0.95\linewidth]{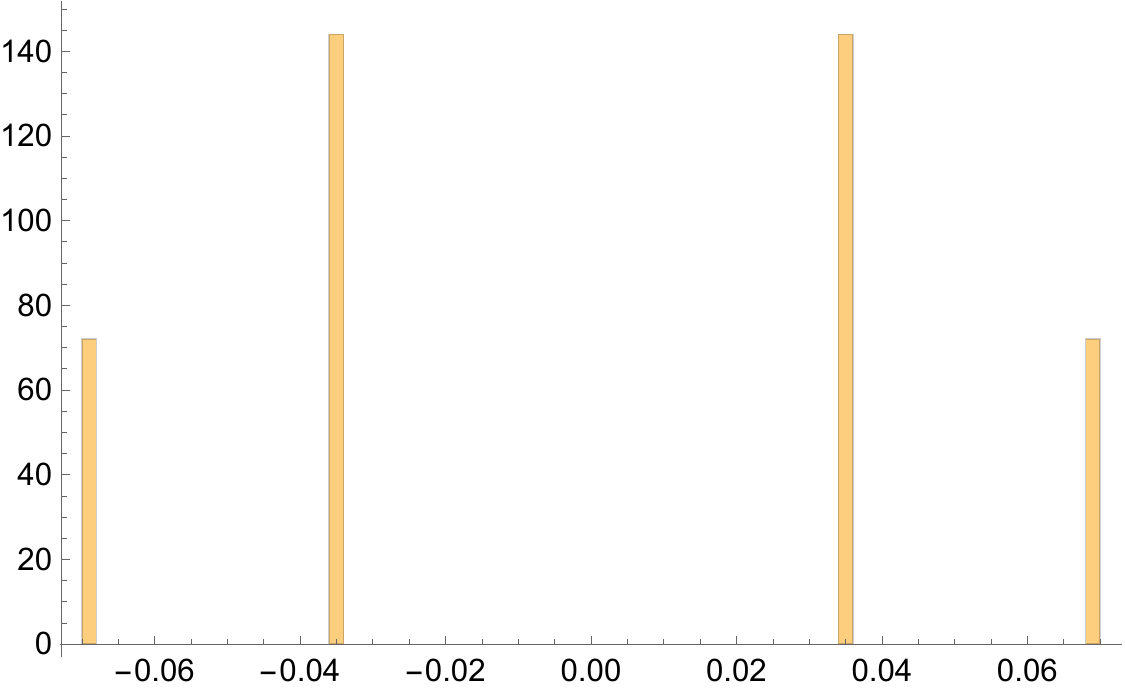}
    \includegraphics[width=0.95\linewidth]{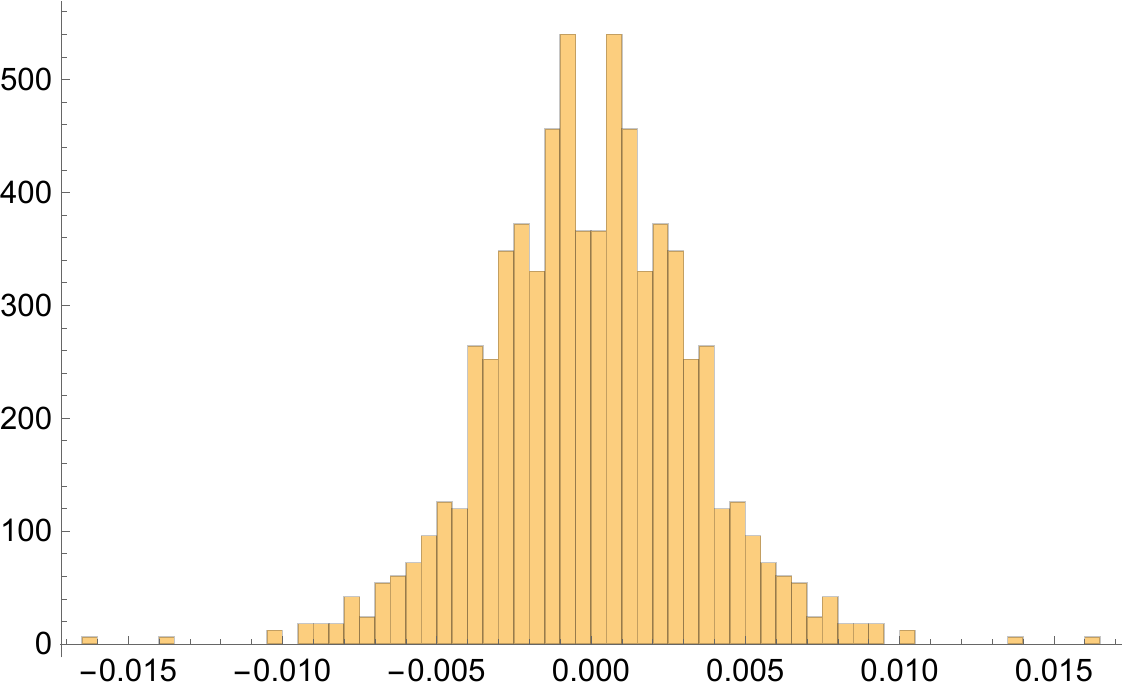}
    \caption{Distribution of monom coefficients, for one of the S-shell  state with $L=0,S=5/2,I=1/2$
    (upper plot) and one of the
    P-shell  state $L=1,S=1/2,I=1/2$) (lower plot). }
    \label{fig_Hist_L1}
\end{figure}

\section{Can  pentaquarks be chaotic?} \label{sec_chaotic}
Quantum many-body (and some few-body) systems are known to
undergo transition into the so called ``quantum chaos" regime, for review see e.g. \cite{Zelevinsky:1996dg}.
This refers to atoms or nuclei, with several particles/holes 
 near the filled shells (magic numbers). As a specific historical example, it was the atom of Cerium, with four valence
 electrons or 12 coordinates, which was shown by \cite{flambaum1997applystatisticallawssmall} 
 that  even the lowest states are chaotic.
 
 Since this phenomenon is generic, one may also expect it to
 take place for multiquark systems. Pentaquarks,
 the subject of this paper,  have 12 Jacobi coordinates, the same number
 as the Cerium atom. So one may expect to 
find similar ``transition to chaos"
in them. (One of us 
\cite{Shuryak:2019zhv} has already discussed 
this issue for baryons, pentaquarks and their admixture
in a particular light-cone model, with  different wave functions and different set of basis states.)
 
 The simplest manifestation of quantum chaos is $random$ (Gaussian-like) Porter-Thomas distribution of the
 amplitudes of the (exact) Hamiltonian eigenstates,  expanded in terms of some natural basis, e.g. of the single-body Hamiltonian. 
In Fig.\ref{fig_Hist_L1} we show two distributions of the ``monom coefficients" for our fixed-energy states.

The upper is for the $S=5/2,I=1/2$ S-shell state which
we obtained both analytically and numerically.
We recall that its wave function consists of three
terms of different structure, times permutations. And the upper histogram confirms that there are only 2 values (up to a sign) of such amplitudes, repeated multiple times.
Furthermore, recall that there are $4!=24$ quark permutations, while occupancy in the histogram turns out to be $72=24*3$. This state clearly is very regular, far from
being chaotic.

The lower histogram is for one state
from the  P-shell $L=1$ pentaquark.  
As one can see from it, the overall shape of the amplitude distribution
is indeed similar to a Gaussian, although deviations from it are also clearly seen. 
The most obvious is
a dip near zero value. Similar dip often is due to ``finite size effects", for example in Dirac quark eigenvalue
spectrum in lattice or instanton studies of the QCD vacuum. However in this case the dip shape is well understood in chiral random matrix model \cite{Shuryak:1992pi,Verbaarschot:1993pm} , in terms of the box size and number of states. The analytically predicted distributions were well confirmed in
multiple lattice studies.
It is not clear to us what exactly creates the
dip in this histogram.  Let us just mention that the
 total number of nonzero coefficients 14400 in a monom basis of size 746496, so it is still rather sparse.

Another feature of the lower histogram are regularly
placed peaks. Perhaps an indication of a quantum system with
the phase space in which both chaotic and regular
motion are present, with the former being dominant.

Quasi-random coefficients of the wave function in a particular basis is of course only one of many observables one can study.
Probability distributions over single-body properties (e.g. spatial distributions or those in single-body-energy) can be compared
to predictions of some statistical models (e.g. those for Fermi gas with certain temperature and entropy).
Indeed, a single many-body state can have  a thermal
description in terms of single-body observables!
We defer such studies to future works.

Summarizing, we presented evidence that transition to
quantum chaos in pentaquarks does happen in the second ($L=1$) shell states.

\section{Nucleon-pentaquark mixing}

\subsection{The ``unquenching" of hadrons}

  The ``Unquenching" of mesons and baryons -- in attempts to quantify the four-quark sector of mesons and the five-quark sector of baryons -- has a long history but is still far from being completed.

  From the 1960's to about a decade ago the experimental and theoretical status 
  of multiquark hadrons was assigned  into the general realm of ``exotica", which looked suspicious 
  as lacking firmly established facts. It has all  changed now.
The revolutionary change perhaps has not yet settled
in the minds of the community at large, so 
 let us illustrate it by  focusing on the pillar of QCD
spectroscopy, heavy quarkonia. 

While bottomonium spectroscopy is fully consistent
with the $\bar b  b$ interpretation, using a simple Cornell potential reproducing the states,
%nicely describing observed  states of S,P,D shells.
already $\bar c c$  charmonia  are different. The first
 ``non-charmonium" state discovered was called $X(3872),J^P=1^-$ state. 
 Originally called
$X,Y,Z$ etc for unknown structure, in the
current particle data group (PDG) terminology they are split into two categories. 
``Axial vectors" (quantum numbers $I=1,J^P=1^+$)
have nonzero isospin and therefore are considered ``pion-added". 
Since the 2024 PDG classification, they are identified as teraquarks and are called  $T_{\bar c c 1}$. Yet for vectors $J^P=1^-$ and pseudoscalars $J^P=0^{-}$ (zero isospin or ``$\sigma$ added"),  the PDG uses still their generic notations, $\psi,\eta_c$ respectively. This is unfortunate, especially  in view of the observations that vectors and axials seem
to form near-degenerate pairs, perhaps being ``chiral partners".

(Some charmed tetraquarks  presumed to be  ``molecular" (deuteron-like)  two-meson state (e.g. $DD^*$) just below the threshold, but not all. Lately many more tetraquark states are found, not only with $\bar c c$ but also $c c$ pair, with even all-charm $\bar c \bar c c c$ tetraquark found at LHC. For the details on such tetraquarks and mixing we refer to~\cite{Ferretti:2013faa} and references therein.)
  
  Completing this detour to charmonia, let us make an important point:  the number of pion-admixed and sigma-admixed known states seems to be equal. With four $T_{\bar c c 1}$ and four 
 ``extra-$\psi$"  states  unfit to be identified as  $\bar c c$ levels, for which we 
  observe a kind of $\pi-\sigma$ symmetry reminiscent of the original chiral symmetry.
  Chiral doubling in heavy-light systems was originally noted in~\cite{Nowak:1992um,Bardeen:1993ae,Nowak:2003ra} and since confirmed 
  experimentally by the BaBar~\cite{BaBar:2003oey} and CLEO collaborations~\cite{CLEO:2003ggt}.
%  (ii) Because of kinematic reasons, similar phenomena
%  in bottomonia $\bar b b $ family are more suppressed.

\begin{table}[t]
    \centering
    \begin{tabular}{|c|c|c|}
    \hline
     state    &  $N\pi$ & $N\sigma $ \\ \hline
    $N^*(1440)\, 1/2^+$ & 0.55-0.75 & 0.11-0.23 \\ 
   $N^*(1520) \,3/2^+$ & 0.55-0.65 &  $<0.1 $ \\
    $N^*(1535) \,1/2^-$ & 0.32-0.52 &  $0.02-0.1 $ \\
     $N^*(1650)\, 1/2^-$ & 0.5-0.7 &  $ 0.02-0.18$ \\
      $N^*(1675) \,5/2^-$ & 0.38-0.42 &  $0.03-0.07 $ \\ 
    $N^*(1675)\, 5/2^-$ & 0.38-0.42 &  $0.03-0.07 $ \\
        $N^*(1680)\, 5/2^+$ & 0.6-0.7 &  $0.09-0.19 $ \\
      $N^*(1700) \,3/2^-$ & 0.07-0.17 &  $0.02-0.14 $ \\
        $N^*(1710)\, 1/2^+$ & 0.05-0.20 &  $<0.16 $ \\
            $N^*(1720)\, 3/2^+$ & 0.08-0.14 &  $0.02-0.14 $ \\  
       $N^*(1875)\, 3/2^-$ & 0.03-0.11 &  $0.02-0.08 $ \\ 
          $N^*(1880)\, 1/2^+$ & 0.03-0.31 &  $0.08-0.40 $ \\ 
             $N^*(1895)\, 1/2^-$ & 0.02-0.18 &  $<0.13 $ \\
                    $N^*(1900)\, 3/2^+$ & 0.01-0.20 &  $0.01-0.07 $ \\ 
            $N^*(2060)\, 5/2^-$ & 0.07-0.12 &  $0.03-0.09 $ \\   
             $N^*(2100) \,1/2^+$ & 0.08-0.32 &  $0.14-0.35 $ \\
              $N^*(2120) \,3/2^-$ & 0.05-0.15 &  $0.04-0.14 $ \\
            $N^*(2190) \,7/2^-$ & 0.1-0.2 &  $0.03-0.09 $ \\
         \hline
    \end{tabular}
    \caption{Branching ratios of $N\pi$ and $N\sigma$ decays of $N^*(mass) J^P$ resonances (left column),  from PDG24}
    \label{tab_branchings}
\end{table}

%%%%%%%%%%%%%%%%
 In general, the appearance of the $\sigma$=$f_0$ meson in hadronic reactions
 has been artificially suppressed over the years. To emphasize our point regarding
 this important chiral mixing,  let us provide another little known
 case of the ($\pi-\sigma$) symmetry.
A list of branching ratios (from PDG24) of main-listing nucleon resonances, into both  decay modes is given in the Table\ref{tab_branchings}. While the pion branchings tend to be larger,
 this is actually a reflection  of the larger phase space 
for $N\pi$ compared to $N\sigma$.    The pertinent couplings,
$g_{NN^*\pi}$ and $g_{NN^*\sigma}$, are basically equal
within errors. This is  what  ``naive" chiral symmetry suggests.
%%%%%%%%%%%%%%%
 
  After this two brief detours to the PDG data, let us turn to 
  the theory. 
 The traditional strategy to evaluate the 5-q admixture to baryons consists of two 
 steps: \\
 (i) formulation of certain ``meson admixture operators" adding $\bar q q$ pair to a baryon, \\ (ii)
 followed by projections of their product
to ``good tetraquark states"  obeying general requirements
such as Fermi statistics.

%%%%%%%%%%%%%%%%

Before discussing  the  technical implementation of steps (i) and (ii) let us mention some key facts about chiral symmetries and their breaking.
%%%%%%%%%%%%%%%%
Long before the discovery of QCD, Nambu-Jona-Lasinio 
\cite{Nambu:1961fr} have introduced hypothetical a 4-fermion Lagrangian, and have shown that for a large enough coupling,  it can break spontaneously the $SU(N_f)$ chiral symmetry.
The  discovery of fermion zero modes of instantons by t' Hooft \cite{tHooft:1976snw}, had shown that QCD does generate non-perturbative multi-fermion $2N_f$ interactions. For two flavors ($N_f=2$) 
the resulting 4-fermion Lagrangian %is often presented %in bosonized form
can be written in many forms, in particular as
\be L_{\rm 'tHooft}=G_{\rm 'tHooft} \big( \sigma^2+\vec \pi^2 - \eta'^2-\vec \delta^2 \big)
\ee
Here the binary quark-antiquark operators are given with respective mesonic names, as
$$  \sigma=\bar q q, \,\, \vec \pi=\bar q \vec \tau i\gamma_5 q,\,  
\eta'=\bar q i\gamma_5 q, \,\vec \delta=\bar q\vec \tau  q$$
 The first two attractive terms are the same as in the hypothetical Nambu-Jona-Lasinio
Lagrangian. Note that the two last  terms have negative (repulsive) sign, making a violation of the $U(1)_a$ chiral symmetry explicit.
As shown in \cite{Shuryak:1981ff} and confirmed since by
instanton phenomenology, the coupling constant $G_{\rm 'tHooft}$ is  large enough to {\em break spontaneously } chiral $SU(2)_a$ symmetry.
Also it was shown in review  of correlation function in the QCD vacuum \cite{Shuryak:1993kg}, that the  't Hooft Lagrangian is
indeed responsible for flavor mixing 
of all scalar and pseudoscalar mesons, making them
remarkably different from others (vector,tensor...)
channels in which very small flavor mixing takes place (``Zweig rule").

%%%%%%%%%%%%%%%%%%
\subsection{Admixture via 't Hooft Lagrangian}
 Since both the  NJL or 't Hooft Lagrangian  can turn one quark into $qq\bar q$,
it seems logical to start with them, as the  ``admixing
operators" for flavor mixing.

Let us start by recalling  that the  't Hooft Lagrangian 
is built out of instanton zero modes. For three flavors,
it is a 6-fermion operator which can be represented as a
flavor determinant.  
For two flavors $u,d$ the 't Hooft operator is 4-fermion one, which
can be written  with spin and isospin indices as follows
\ba
\epsilon_{f1 f2}\epsilon_{g1 g2}\bigg((1-{1 \over 2N_c})(\bar q_{L f1} q_{R g1})(\bar q_{L f2} q_{R g2}) \nonumber \\
-{1\over 8 Nc} (\bar q_{L f1} \sigma_{\mu\nu} q_{R g1})(\bar q_{L f2} \sigma_{\mu\nu} q_{R g2})
\bigg)
\ea
where $L,R =(1\pm \gamma_5)/2$ refer to the projected left-right spinors, and  $\sigma_{\mu\nu}=(\gamma_\mu \gamma_\nu-\gamma_\nu \gamma_\mu)/2$. The two epsilon symbols yield a determinant. Alternatively  they
can be substituted by flavor matrices via the Fierz-like identity
$$ \epsilon_{f1 f2}\epsilon_{g1 g2} =(\hat 1-\vec \tau_{f1 g1}\vec \tau_{f2 g2})/2 $$

The $2N_f$ entries in the 't Hooft vertex  originate
from a single instanton. Hence,  
the distance between them
 is restricted by the mean instanton size~\cite{Shuryak:1981ff} to $$r_\sigma\sim\rho \approx 1/3\, fm$$
This size is comparable to the radii of the scalar $\sigma$  and pseudo-scalar  $\pi$
mesons, and defines many of their internal properties. 
 However, the sizes of baryons and tetraquarks are
 much larger. This drastically
 reduces the effect of the first-order contribution of the 't Hooft Lagrangian, as we now show.

Let the 4 fermion coordinates (to be defined below) be $\vec X_i, i=1,2,3,4$. In coordinate representation , the t' Hooft 4-fermion
operator is proportional to the product of 4 zero modes
\be \label{eqn_Hooft}
T_{\rm tHooft}\sim \prod 
\gamma_\mu\partial_\mu\phi_0(\vec X_i-\vec y)
\ee
where $\vec y$ is the instanton center and 
the (full) zero mode is
\ba
\phi_0^{a \nu} &=& \psi_0 {(1-\gamma_5)\over 2}{x_\mu \gamma_\mu \over \sqrt{x^2}}U_{ab}\epsilon_{\nu b} \nonumber \\
\psi_0 &=& {\rho \over \pi}{1 \over (x^2+\rho^2)^{3/2}}
\ea
with $U$ being instanton orientation matrix in the $SU(3)$ color space. 

(To understand how momenta in (\ref{eqn_Hooft}) are  related, one has to 
express the zero modes via their Fourier transform $\tilde\phi_0(\vec p)$
\be \phi_0(\vec X-y) \rightarrow \int d^3 p\, \gamma_\mu p_\mu\,\tilde\phi_0(\vec p) e^{-i \vec p (\vec X-\vec y)}
\ee
The integral over the instanton center $d\vec y $ produces overall
 momentum conservation, in the form
$ \delta(\sum_i\vec p_i) $. The expressions
for pairwise product of $\tilde\phi_0(\vec p)$ are well known, and yield  ``instanton form-factors" in many applications, but we will not need them here.)

\begin{figure}
    \centering
\includegraphics[width=0.75\linewidth]{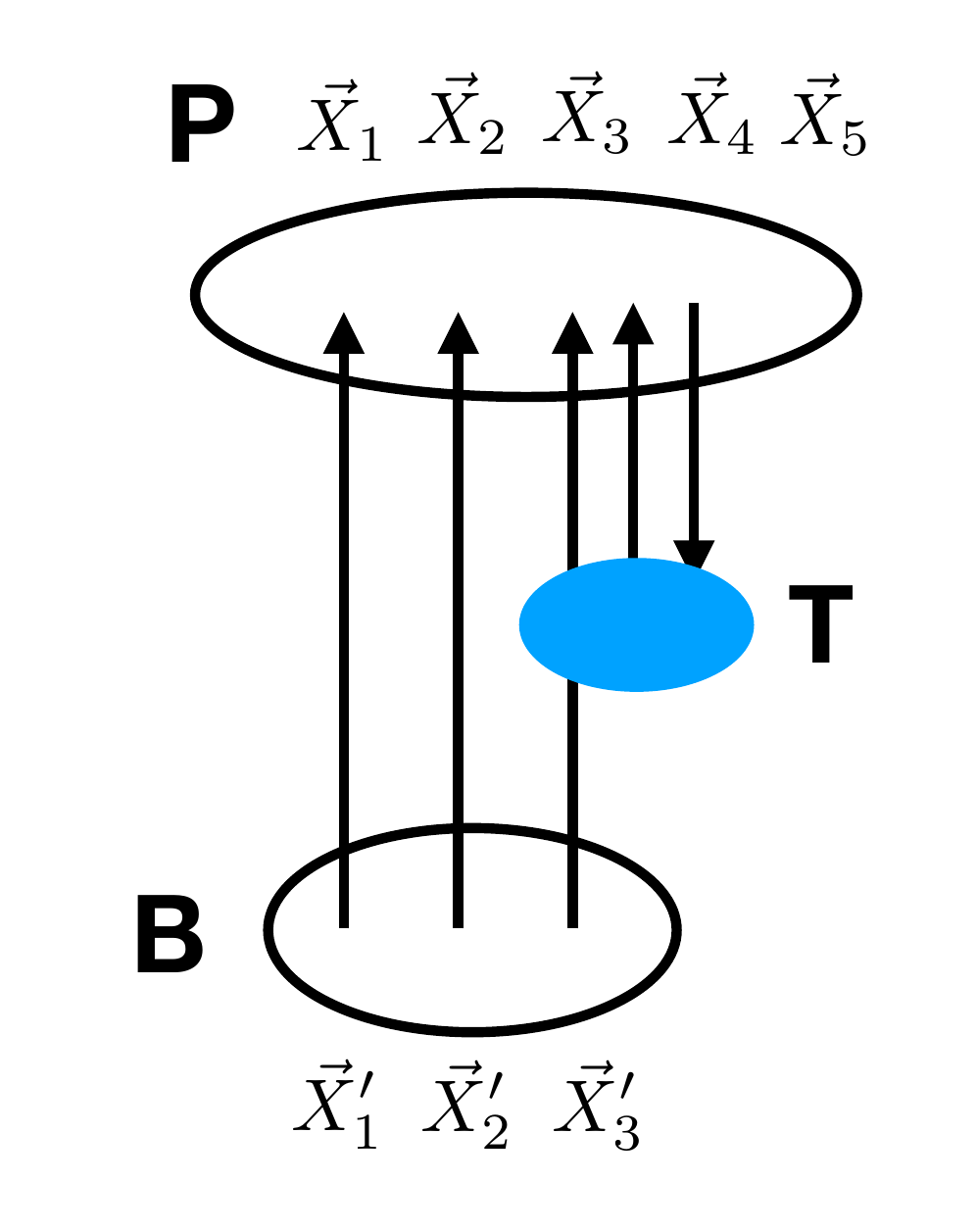}
    \caption{Pentaquark-baryon-meson
    overlap structure, with  the defnition of the relevant in-out coordinates for the quarks.}
    \label{fig_Hooft}
\end{figure}

The notations for the coordinates
in the projection integral $\langle P | T | B  \rangle$ are  explained in Fig.\ref{fig_Hooft}. The pentaquark wave function depends on five coordinates $P(\vec X_i), i=1..5$ while that of the baryon on three $B(\vec X'_n),n=1,2,3$. Let the transition operator describes the  interaction of quark number 
labeled by 3 with a quark-antiquark pair labeled  by 4-5,
\begin{widetext}
\be T_{'t\,Hooft} \sim \int (\prod d^3 X_i)
(\prod d^3 X'_i) \delta(\vec X_1-\vec X'_1) \delta(\vec X_2-\vec X'_2) P(\vec X) B(\vec X')\, T 
\ee 
\be
T=\int d\vec y \,\psi_0(\vec X'_3-\vec y)
\psi_0(\vec X_3-\vec y) \psi_0(\vec X_4-\vec y) \psi_0(\vec X_5-\vec y)
\ee
\end{widetext}

The 4-point function $T$ can be estimated by the standard expansion-of-exponent method
$$\psi_0(\vec X)\sim \rm exp\bigg(-{3\over 2\rho^2}\vec X^2 \bigg)$$
and, after integration over $\vec y$,
we obtain  the vertex function in a Gaussian form
\be T\sim {\rm exp}\bigg( -\frac 9{8\rho^2}(\vec X_1^2+\vec X_2^2+\vec X_3^2+\vec X_4^2)+{3 \over 4}\sum_{i>j}(\vec X_i \vec X_j)\bigg)
\ee 
More simplifications follow from the ``quasi-local" approximation, using  the fact that instanton size is small $\rho \ll |X_i|$. If
we assume  that all 4 coordinates 
in the 't Hooft vertex coincide 
$$\vec X_3'=\vec X_3=\vec X_4=\vec X_5$$
then the three Jacobi coordinates should vanish
$$\vec \beta=\vec \gamma=\vec \delta=0$$
leaving only one nonzero $\vec \alpha\neq 0$.
Therefore, the overlap integral should scale as a large power of the small parameter
\be \langle P | T | B  \rangle \sim \bigg({\rho \over R}\bigg)^9\ee
(where $R$ is the pentaquark size)
and therefore is very small. We conclude that the 
mixing induced by the first order in 't Hooft quasi-local operator
is  too small, and is unable to compete with the nonlocal
production of a quark pair in the $\sigma,\vec\pi$
states (viewed as a coherent iteration of the 't Hooft interaction in the QCD instanton vacuum) as described below.

%%%%%%%%%%%%%%%%%%%%%%%%%%%%%%%%%%%%%%%%%%%%%%%%%%%%%%%%%%%%%%%%%%%%
\subsection{``Soft additions" of the $\bar q q$ pairs}
Mixing of states with the same global quantum numbers is a phenomenon well known all over
atomic and nuclear physics. The  old-fashion perturbation theory to first order
\ba \label{eqn_admixture}
\psi(B)&=& N \large[|B \rangle +\Delta \psi \large ] \\
\Delta \psi &=&
\sum_n  {  \langle P_n^* | \tilde T | B \rangle \over M_{P_n}-M(\bar q q)-M(B) } | P_n\rangle \nonumber 
\ea
tells us that the
matrix element of the mixing operator $\tilde T$ gets measured against the energy splitting between the in-out states.  Therefore,  in some cases mixing can be substantial even if $both$ numerators and denominators are  small. Perhaps the best known effect of this type in atoms and nuclei, is the mixing of opposite parity states due to neutral currents in weak
interactions.

The spontaneous breaking of  chiral symmetry by the NJL or 't Hooft interactions is due to high-order bubble chains of $\bar q q$ ``exploding"
in the instanton vacuum, reaching  infinite distance for massless pions
and producing nonzero VEV $\langle \bar q q\rangle \neq 0$.  The 4-fermion t'Hooft vertex (shown in Fig.\ref{fig_Hooft} by a blue ellipse) 
gets split into two $$\sigma^2+\vec \pi^2\rightarrow \sigma(x)\times \sigma(y) +\vec \pi(x)\times\vec \pi(y)$$
chirally symmetric contributions.
The sigma insertion gets disjoint even for arbitrarily well-separated points $x,y$, with the same quark
condensate everywhere. The pion
insertion yields the pion propagator,
or a Yukawa-type potential between
the coordinates $\vec X_3$ and the mean of the coordinates $(\vec X_4+\vec X_5)/2$ in the notations of Fig.\ref{fig_Hooft}. The technical
details on how this happens in the instanton vacuum can be found e.g. in the reviews \cite{Schafer:1996wv, Nowak:1996aj} and references therein.
 
 In summary, we will use the operators of the ``soft $\bar q q$ admixture", a sum of the  
  ``sigma"-scalar with vacuum quantum numbers for the pair, and 
  the ``pion"-pseudo-scalar for the pair. Each $\bar q q$
  vertex is near-local, with the instanton-induced size.
  Their implementation  will be carried below.

%%%%%%%%%%%%%%%

Before we detail our analysis, let us comment on the historic development and the relation between them.  $SU(N_f)$
 chiral symmetry implies that each state has a  ``chiral copy". This includes the nucleon
 state and its chiral copies induced by the unitary rotation  $exp\big(\sigma+i\gamma_5 (\vec\tau \vec\pi)\big) $. Even after its spontaneous breaking (adding nonzero VEV to $\sigma$),
this transformation leads multiple chiral relations
connecting the sigma and pion couplings, e.g. the famed
Goldberger-Treiman relation
$g_A m_N=g_{\pi NN}f_\pi  $. 

 The admixture of a $\bar q q$ pair with vacuum (sigma) quantum numbers referred to as the  $^3P_0$  (S=1,L=1,J=0) model,  was suggested in the 1970's (e.g. \cite{LeYaouanc:1972vsx}), and  actively discussed in the 1990's. In particular,  Isgur and collaborators \cite{Geiger:1991qe} started quantitative studies  aimed at ``unquenching" the nucleon, mostly by adding strange $\bar s s$
 and charm $\bar c c$ components. They also raised the question of whether a large $\bar q q$ admixture is compatible with some  of the successes of the $qqq$ models, such as the values of their magnetic moments.

 The same mixing operator was used in subsequent ``baryon unquenching" works, such as \cite{Bijker:2009up}. They also  used the same baryon-meson 
set of states in the oscillator basis, following on the work in~\cite{Geiger:1991qe}. While  these states do not really represent pentaquarks, they allow for a simple enforcement of Fermi statistics.

Another direction of research focused on
the ``pion cloud" of the nucleons. It is based on  chiral dynamics, with pertinent Lagrangians and diagrams. The pion-based 
admixture is discussed mainly in connection to
the issue of {\em flavor asymmetry of the antiquark sea}, $\bar u(x) \neq \bar d(x)$. A particularly popular approach defines the admixture  as $N\pi,\Delta \pi$
states in the nucleon, with the pion possessing certain momentum.

It has been pointed out by Garvey \cite{Garvey:2010fi}, that  the magnitude of the flavor asymmetry in this simple model is $equal$ to another puzzling
observable, the magnitude of the {\em quark orbital momentum} (QOM) inside the nucleon. While this model have not
yet described these two observables quantitatively,
its predictions are of the right magnitude.
These works affirmed the general conclusion  that the nucleon's 5-q sector 
is $not$ small.

The issue of quark orbital motion inside the nucleon has 
been address from a different point of view  in our recent work
\cite{Miesch:2025vas}, where we studied the possible admixture of the 
second D-shell ($J^P=2^+$ ) $N^*$ resonances.  However, we concluded  that such admixture to the nucleon should remain small, even if the tensor forces are large.

\subsection{Sigma-induced admixture of pentaquarks to baryons}
As  noted long ago, 
  the scalar quantum number of a quark-antiquark pair
  requires spin $S=1$ and orbital momentum
 $L=1$ added together to $J^P=0^+$. This mechanism
 is believed to be the leading reason for the observed quark orbital motion in the nucleons. It is this important physical observation supplemented by the strictures of chiral symmetry and Fermi statistics, that have motivated our current  P-wave analysis  of the pentaquark states.

% Thus the
% admixture
%operator should be proportional to $ T_{scalar}\sim (\vec %S \vec L)$
%spin-orbit combination.

Multiple studies of ``baryon unquenching" (e.g. \cite{Bijker:2009up} have
described the admixed 5-q states  as a ``baryon plus a meson" state. 
These states were described as the  $N+\pi$ or $\Delta+\pi$
states, with rather large pion momenta. While  the search for baryon resonances are 
done this  way, and while a loop diagram with a
virtual pion does lead to some flavor asymmetry of the antiquark sea, we do not
see why one should neglect e.g. $N+\sigma$ channels (see below). More generally, why not 
look at all pertinent pentaquark states.

Let us start with the sigma-induced admixture operator 
studied in many works since the 1970's. Its form in momentum space is~\cite{Bijker:2009up} 
\bea \label{eqn_vacuum_operator}
&&T =-3\gamma_0 \int d\vec p_4 d\vec p_5 \delta(\vec p_4+\vec p_5)\nonumber\\
&&\times e^{-r_\sigma^2(\vec p_4-\vec p_5)^2/6}\, Y_{1}(\vec p_4-\vec p_5)\,\chi_{45}^{color,flavor,spin}\nonumber\\
\eea
Note that  $\vec p_1,\vec p_2,\vec p_3$ refer to the quark momenta in a baryon, and
$\vec p_4,\vec p_5$ those in the scalar  $q\bar q$ pair, or Jacobi coordinates $\vec \alpha,\vec \beta$ and $\vec\delta$, respectively.
 The parameter  $r_T$ is the radius of the region inside the baryon, where the
vacuum  $q\bar q $ pair appears, with $r_T\sim 0.3\,\rm fm$. This value is fully consistent
with its instanton-induced origin, and the typical size
of the instantons in instanton liquid model.

We will use the coordinate version of (\ref{eqn_vacuum_operator}) with minor changes
(e.g. the exponent reads $e^{-(\vec X_4-\vec X_5)^2/r_\sigma^2/2}$).
The color-flavor parts of the wave function are $\chi_{45}^{color,flavor,spin} \sim \delta^{f1 f2} \epsilon_{c4,c5,c6} $. The coordinate and spin must be convoluted to produce a
total angular momentum $J_{45}=0$ due to the scalar operator $\vec S \cdot \vec L$
in the operator $T$. The emission of a sigma is tagged to the 3-quark, as 
our pentaquark wave functions are all properly anti-symmetrized.   
Those are $projected$ on the product of the 
vacuum pair production operator $T$ (\ref{eqn_vacuum_operator}), and the baryon wave function $B$
\be A_{mix}^n=\langle P_n^* | T | B \rangle \ee
 Note that only pentaquarks  with the first power of coordinate $\vec\delta$ in their wave function contribute, since the vacuum operator $T$  is linear in 
 the first angular harmonics of this vector. Thus, the orbital admixture in baryons arises
from their mixture to pentaquarks.

Let us start with the radial integrals first, and determine how many ``pentaquark shells" may contribute to this projection. Note that since we
have appropriately anti-symmetrized the pentaquarks, there is no need to use cumbersome
and very restrictive oscillator basis to do it, as done in the literature.
We define the radial part as
\bea \label{eqn_mix_radial}
&&A_{\rm mix}^n(radial) =\int \alpha^2 d\alpha \beta^2 d\beta  \gamma^2 d\gamma \delta^2 d\delta   \nonumber \\
&&\times R_n(\sqrt{ \alpha^2+ \beta^2 +\gamma^2+ \delta^2}) \tilde T(\delta) B(\sqrt{ \alpha^2+ \beta^2 })\nonumber\\
\ea
where $\tilde T $ is the Fourier transform of the vacuum pair operator $T$, 
$R_n$ are the respective solutions
of the radial equation in 12-d space, and $B$ is the ground state baryon solution in 6-d.
The results for the radial modes $n=1..5$ normalized to the first one are as follows: % from penta_bar_mes.nb 
\be \label{eqn_Amix}
A_{\rm mix,1..5}(radial) =1,0.044, 0.203, 0.0340, 0.075
\ee
suggesting that the sum should converge relatively well.
Note that the overlaps for odd $n$ -- those with odd number of nodes in the wave functions -- are smaller than those with even $n$. In practice, perhaps keeping $n=1$ and $n=3$ would
be enough.
%these overlap integrals are relatively small already for the first state,
%since it has a radius (in hyperdistance) of about 2 fm, while the baryon size is still
%of 1 fm. Note also that  

The energy splittings or gaps for $n=1..5$ from the radial Schrodinger equation, 
are listed in  (\ref{eqn_gaps}).
%\be Gap_n=M_{penta,1..5}(radial)-M_{penta,1} =(0, 0.32, 0.62, 0.89, 1.16) [GeV]
%\ee
While they may appear to be small compared to 
the overall mass scale $M\sim 2\, GeV$,
 the denominator of the first order perturbation 
$$ \sum {A_{\rm mix,n} \over Gap_n+M-M_n -m_\sigma}$$
is actually $M-M_n -m_\sigma\approx 0.5\,GeV$ rather than just $M$. Therefore
Isgur's ``completeness approximation"
in which the denominator dependence on $n$ is neglected,  does not
hold well. These gaps help in the
convergence of the sum. With gaps included, (\ref{eqn_Amix}) changes to
$$
1., 0.040, 0.173, 0.027, 0.056$$ 
(We used a sigma meson of $M(\bar q q)\approx 0.5 \, \rm GeV$,
and for the baryon mass we used the spin-average $N-\Delta$ mass $M(B)\approx 1\, \rm GeV$. Note that
for pentaquarks we also do not include the spin-dependent forces). 

To complete the task, on has to calculate the overlaps of the coordinate-color-flavor-spin
wave functions.
Finally, the actual magnitude of the pentaquarks admixture is proportional to the overall parameter $\gamma_0$ in the vacuum production operator $T$. While its magnitude can be estimated using the instanton liquid parameters, below we choose to fit empirically to the observed nucleon ``sea".

\subsection{Pion-induced admixture to nucleons}
The motivation for adding a "pion-induced admixture"
(also known as a ``pion cloud") has been detailed above, 
so we now proceed to its implementation.

 One important point is change of the spatial parity
 in the transition operator.
 While the sigma-induced operator $T_\sigma$
includes the spin-orbit $(\vec S \vec L)$  combination
with positive parity $P=+1$,
the pion-induced one includes a different  spin-momentum combination
\be \label{eqn_SP}
O_{Sp}=(\vec S\vec P) \ee  with negative parity $P=-1$. Note that  the pion (unlike the sigma) is a  Nambu-Goldstone mode, so its interaction with any hadron must
vanish in the long wavelength $\vec P \rightarrow \vec 0 $ limit, as is the case here.

A standard way of implementing this, is to
use the momentum representation of the wave functions for fixed $\vec p$,
evaluate the pertinent overlap, and then Fourier back to
coordinate space to make use of the pentaquark wave functions. This way is straightforward,
but very cumbersome. 

Here we follow a different approach, whereby 
$\hat{ \vec P} =-i {\partial/\partial \vec x }$ in the coordinate representation.
The operator (\ref{eqn_SP}) in the coordinate representation, is made first to act on
the pentaquark side. For $L=1$ it is a complicated combination of orbit-color-spin-isospin  prefactors 
\be F_{n,j}=\vec a_{n,j} \vec \alpha+\vec b_{n,j}\vec \beta +\vec c_{n,j} \vec \gamma +\vec d_{n,j} \vec \delta \ee 
linear in Jacobi coordinates $\vec \alpha...$. The indices $n$ label  the 24
pertinent pentaquark states,  while the
index $j$ runs over  the pertinent monoms, about  $O(10^4)$ 
of them. (We note that acting by a derivative on a multi-component "sparse array" of functions $F_{n,j}$ is straightforward in Mathematica.)

The derivative also acts on the radial WF $R(Y)$, a function of
 the hyper-distance $Y$.  This leads to a combination of the form
\be
\label{12TERM}
 \bigg(\vec S{\partial  F_{n,j} \over \partial \vec \delta }\bigg)  R(Y)+  \bigg(\vec S{\vec  \delta  \over Y }\bigg)  F_{n,j} R'(Y)  
\ee
in which each contribution picks terms 
containing only the 4-th Jacobi vector $\vec \delta$ in $ F$. After averaging over the directions of $\vec \delta $,  
both terms are proportional to $d_{n,j}$ (which is 
 a multidimensional tensor in color-spin-flavor indices, different for 
 each of the states $n$. It is then 
multiplied by  $\vec S=\vec S_4+\vec S_5$  spin operator, and
projected  to the product of baryon-meson
wave functions,  over all color-spin-flavor indices.

The second term in (\ref{12TERM}) (with a derivative $R'$ of radial pentaquark WFs) 
picks  a coefficient $1/12$. Evaluating  the
integral over $\alpha,\beta,\gamma,\delta$ as in  (\ref{eqn_mix_radial}), we find that
its ratio to the first term is only
$\approx -0.054$, a relatively small
correction. %from penta_bar_mes.nb

We have evaluated the projected coefficients or ``mixture amplitudes"  $C_n$
in the proton wavefunction shift $\Delta\psi_\pi$ in (\ref{eqn_admixture}),
for all $n=1..24$ and $L=1$  pentaquark states. The results 
for the $S=I=1/2$ are listed in Table~\ref{tab:my_label}.
Note  that the pentaquark states should 
have the  proton $S=I=1/2$ quantum numbers, thus the baryon-meson
isospin can  be either $p\pi^0$ or $n \pi^+$.

\begin{table}[h!]
    \centering
    \begin{tabular}{|c|c|c|c|c|c|c|}
    \hline
     $C_n$   & 1 & 2 & 3 & 4 & 5 & 6 \\ 
       &  -.0998 & .1230 & -.0177 & .2269 & .1067 & -.0829 \\ \hline
                  7  & 8 & 9 & 10 & 11 & 12 & 13\\
         -.2159 & .1099 & .0410 & .0328 & .0763 & .0036 & .0700 \\ \hline
       1'  & 2' & 3' & 4' & 5' & 6' &7' \\
-.00546 & -.1373 & -.2893 & .3924 & .2756 & .2522 & .039740 \\ \hline
 8' & 9' & 10' & 11' &&\\
 .0145 & .1070 & -.0288 & -.00546 &&& \\ \hline
    \end{tabular}
    \caption{Projection coefficients $C_n$ for 13 $S=I=1/2$ and 11 $S=1/2,I=3/2$ pentaquark states with $L=1$, the same as Fig.\ref{fig_Hist_OpPi}.}
    \label{tab:my_label}
\end{table}

\begin{figure}[t!]
    \centering
\includegraphics[width=0.95\linewidth]{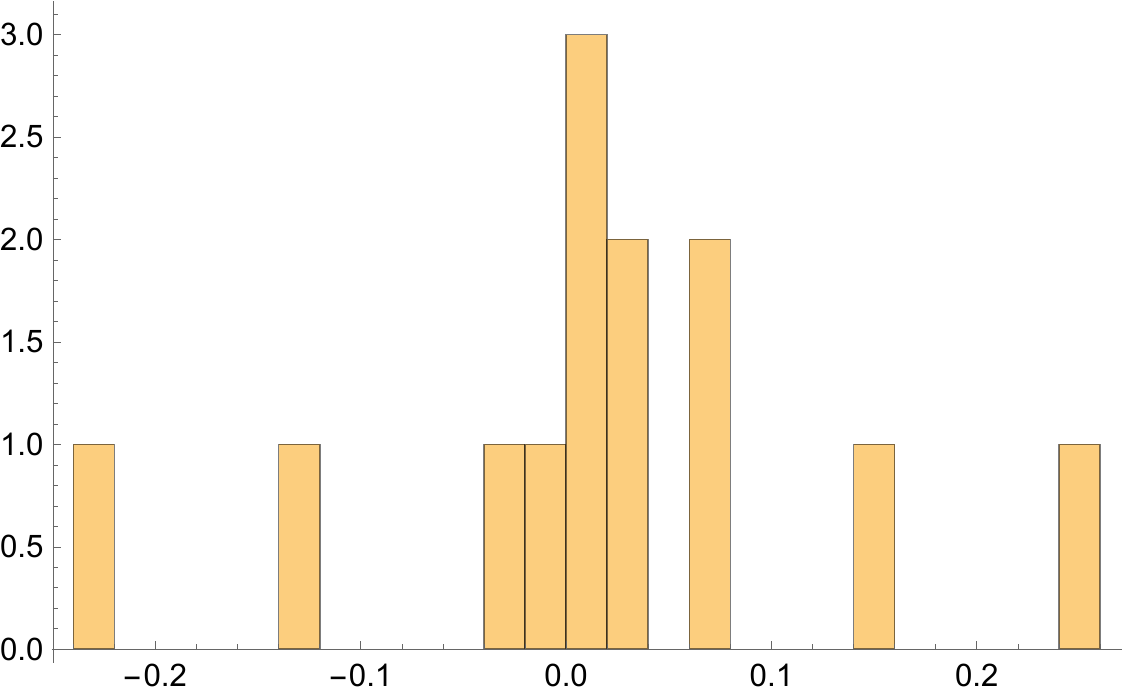}
\includegraphics[width=0.95\linewidth]{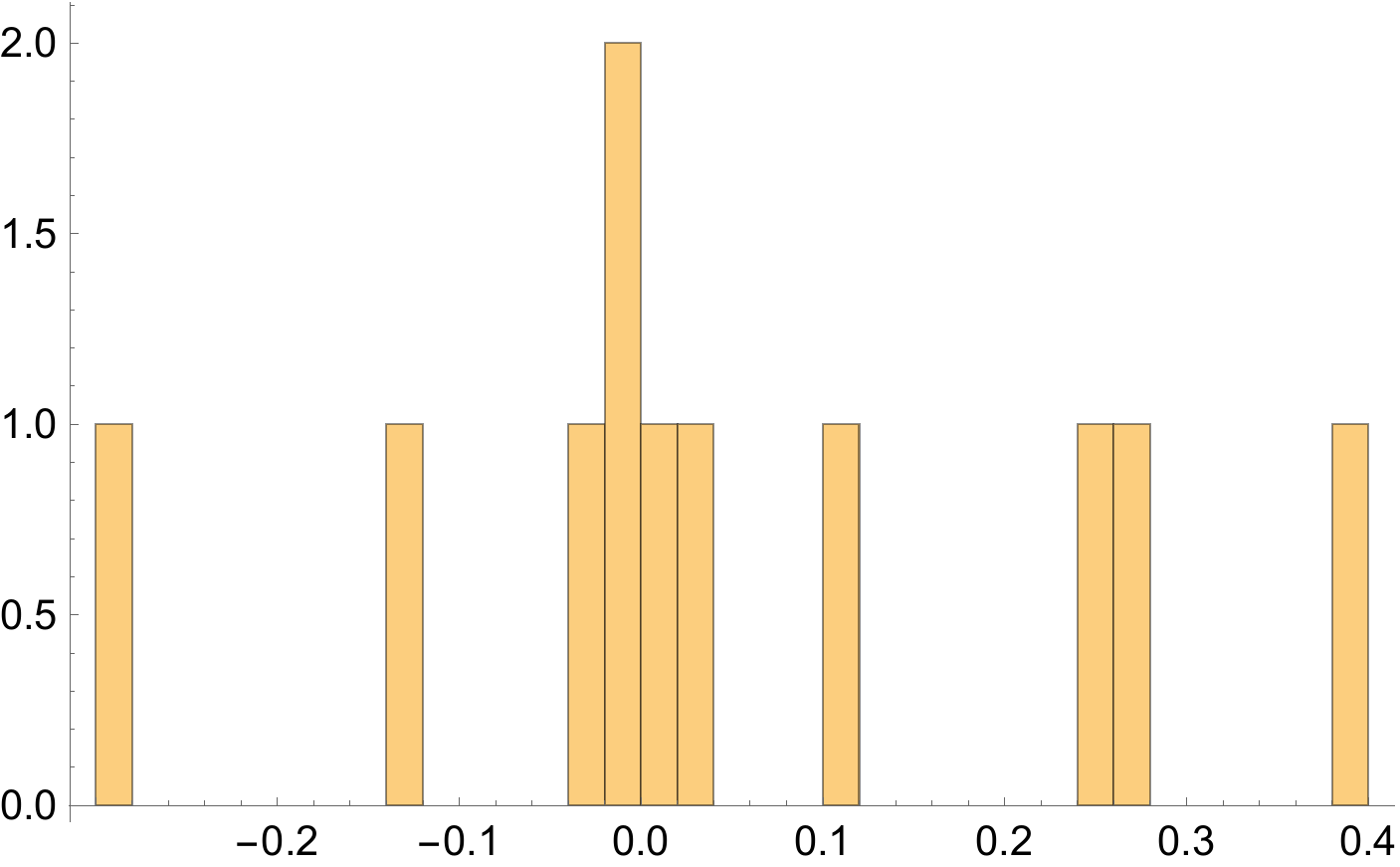}
    \caption{Distribution of pion-induced $\Delta \psi_\pi$ coefficients of pentaquark states
 to a nucleon, from 13 $S=I=1/2$ (upper) and 11 $S=1/2,I=3/2$ (lower) pentaquark
 states from the P-shell ($L=1$)}. 
    \label{fig_Hist_OpPi}
\end{figure}

Although the pion-projected admixture wave function $\Delta \psi_{\pi}$ in (\ref{eqn_admixture}) is evaluated using the pentaquark P-shell basis, there are  qualitative differences between the dependence on the labels $n$ and $j$. The former is by no means a random set,
as is shown in Fig.\ref{fig_Hist_OpPi}, while the states themselves seems to be quasi-random, see Fig.\ref{fig_Hist_L1}.

In Fig.\ref{fig_Hist_OpPi} we show the distribution of the admixture amplitudes 
to the nucleon in $\Delta\psi_\pi$, from a set
of pentaquark states. (These are the same 13 $S=1/2,I=1/2$ states with energies shown in Fig.\ref{fig_pentas_L1}.) What this histogram shows is  far from random, with four states contributing more than the remaining 9. (Note that
the admixture probability is the amplitude $squared$.)

While we have not investigated why this happens, one
can perhaps still use this feature to simplify the
model of $\bar q q$ admixture by keeping only few states.

In summary, while the pentaquark states may be
rather chaotic by themselves (see section \ref{sec_chaotic}), the 5-q $\Delta \psi$ admixture to baryons is clearly   $not$ chaotic, with few states $\sim O(4)$ dominating
all 24  states in  the P-shell.

\begin{figure}
    \centering    \includegraphics[width=0.95\linewidth]{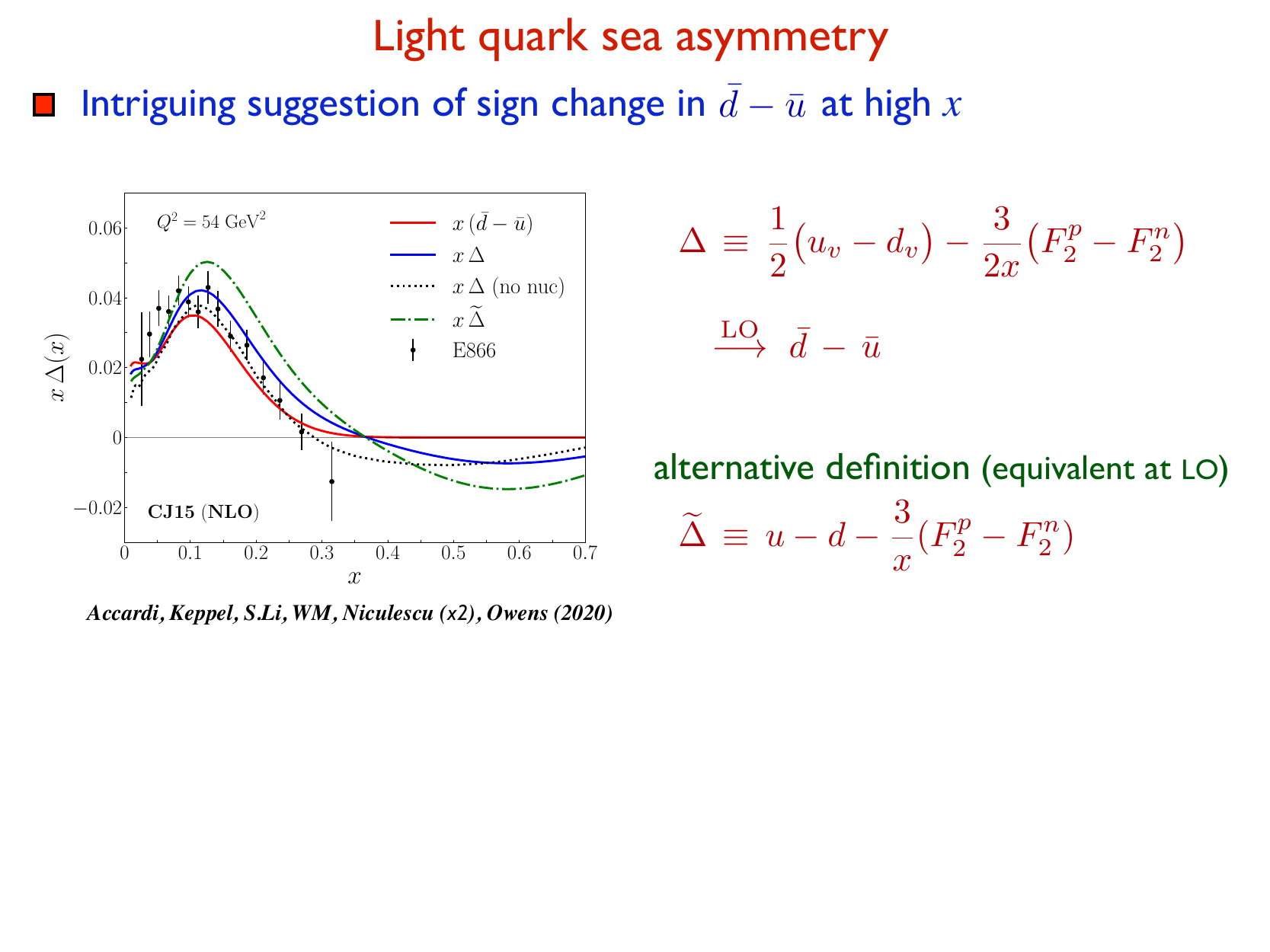}
    \caption{Flavor asymmetry of the light anti-quark sea, from \protect\cite{Accardi:2019ofk}. Experimental points are from E866 experiment, and they are compared to parameterization
   of the combination $x(\bar d-\bar u)$ shown by the red curve. 
   (Ignore other curves.)
   }
    \label{fig_flavorasymmetry}
\end{figure}

%\cite{Diakonov:2012yi}
\section{Observables} \label{sec_observables}
By now, we have defined a $unique$ WF in the 5-q sector
admixed to the nucleon
\be \Delta \psi=\sum_n C_n |P_n\rangle \ee
both for the sigma and pion-induced operators.
Yet the total physical result is proportional to the 
triple couplings $g_{NP}^{\sigma,\pi}$ which are  so far left
undefined, but can be fitted to pertinent observables.  

The most accurately measured one is the
flavor asymmetry of the ``quark sea". Let us remind
the reader how this is obtained. In Fig.\ref{fig_flavorasymmetry} we show the data
from  E866 experiment \cite{NuSea:2001idv} for $x(\bar d-\bar u)$ parton distributions, with some parameterizations of the structure functions. From this we can deduce a 
difference in the number of antiquarks 
\be \label{eqn_isospin_asymmetry}
\int_0^1 dx (\bar d -\bar u) = 0.118\pm 0.012 \ee
with most error stemming from small $x$. 

(Note that the total momentum carried by this admixture, given by the integral under the curve in this figure, 
\be \int_0^1 dx x(\bar d -\bar u) \approx 0.006
\ee
may appear to be more accurate. But, as seen in Fig.\ref{fig_flavorasymmetry}, 
it is unreliable since this integral
can be changed by $x>1/3$ region in which
no actual measurements were made.)

 Using our results for the weighted 5q admixture $\Delta\psi$ wave function, 
 we can calculate the antiquark asymmetry
\be {\langle \Delta\psi \tau^3_{\bar q} \Delta\psi\rangle 
\over \langle \Delta\psi  \Delta\psi\rangle} \approx 0.335 \ee
%{\int d\alpha d\beta d\gamma \langle \delta \psi \tau^3_{\bar q} \delta \psi \rangle \over \int d\alpha d\beta d\gamma d\delta |\langle |\delta \psi|^2 \rangle}\approx \ee
If the probability of the 5q configuration
in the nucleon is $P_{5q}$, then 
the weighted isospin asymmetry is
\be \langle \bar d-\bar u\rangle =0.335 {P_{5q} \over 1+P_{5q}} \ee
Comparing this to the experimental value 
(\ref{eqn_isospin_asymmetry}),  we find 
 $P_{5q}\sim 0.5$.

The contribution to the axial charge of the nucleon $g_A$ is related to the operator $\langle \sum_{j=1..5}\sigma_3(j) \tau_3(j) \rangle$, for which the mean value over $\Delta\psi$
is  0.249 in our analysis. A comparison to the
accurately known experimental value
\be g_A=1.267={5/3+0.249*P_{5q} \over 1+P_{5q}}
\ee
yields $P_{5q}\approx 0.4$. 

As  discussed in the Introduction, the second 
major observable is the orbital motion inside the nucleon. The 5q part carry $L=1$ by construction, and the 3q part $L=0$. The spin sum rule requirement  yields also $P_{5q}\approx 0.4 $.

Finally, there are of course the {\em magnetic moments} of the proton, neutron and hyperons, as well as proton and neutron
form-factors measured both in elastic channels and in
transitions to particular resonances.
We expect to
discuss those in future works, in which the pentaquark theory and
its admixture will be taken to the light front formulation.

\section{Summary and discussion}
In our study of the light quark pentaquarks
we used novel methods in the construction of the  representations of the permutation groups,
in this case of $four$ quarks $S_4$. Using a ``brute
force" approach based on tensor product of permutation generators in
orbital-color-spin-flavor spaces, we were able to construct
fully antisymmetric wave functions of the $L=0$ and $L=1$ pentaquarks, explicitly in the monom space of dimension  $ d_{monoms}=3^6 \times 2^5 \times 2^5=746496 $. The actual number
of nonzero terms in the $L=1$ shell states is of order $O(10^4)$. Note that the pertinent amplitudes are in this case
not numbers but functions of 12 coordinates. Still, with Wolfram  
Mathematica, one can work with those straightforwardly, e.g. using various spin and differential operators in their traditional notations. For some $L=0$ pentaquark states we also were
able to obtain wave functions analytically,
by more standard 
Young tableaux  representations. Needless to say, we have checked
that results of both approaches do agree.

We presented evidences that the P-shell $L=1$ pentaquark states 
are already displaying some evidences for the ``quantum chaos" phenomenon, namely a near-Gaussian coefficient distribution. It is amusing that in the hadronic world we find it first
for the pentaquarks with a quark core of 4 and 12 Jacobi coordinates, the same
number as for a Cerium atom with four valence electrons~\cite{flambaum1997applystatisticallawssmall}.

%\section{Summary and discussion}

In the Introduction, and in particular in Fig.\ref{fig_bridge}, we have illustrated our main goal, ``to bridge" hadronic spectroscopy with partonic observables, by constructing an intermediate ``unquenched" description of hadrons at a resolution $Q^2\sim 1 \, GeV^2$. 
It is supposed to complement the classic description of light baryons as $qqq$
states by an additional $\bar q q$ pair, described as admixing with pentaquarks. 

The distinction between charmonia
$\bar c c$ and tetraquarks
$\bar c c \bar q q$, has been developed in the
last decade, with at least 4 tetraquarks 
in which $\bar q q=\pi$ and other 4 in which
$\bar q q=\sigma$. Multiple hadronic transitions between charmonia and tetraquarks
are observed, providing pertinent information about their admixtures.

The problem we studied in this paper, the mixing between light quark
 baryons and pentaquarks, is similar in spirit. Admittedly, it is much
 more complicated and 
  much less developed. Yet we were able to 
  surpass the technical difficulties, and derive the {\em unique}
  5-q part of the nucleon wave function.

 Our use of the pentaquark states is done ``in bulk", meaning that the wave functions
for these states are used as a (reasonably complete)  basis for admixture calculation. 
 We argued that the admixture operators $T$ should include  both of the $\sigma$ and the $\pi$ $\bar q q$ operators, as it is  the case  for $\bar c c \bar q q$ tetraquarks.
 %(Our estimates for the first-order 't Hooft operator shows that this admixture mechanism should be much less efficient.) 

  The information
 about 5-q component of the baryons come
 mostly from features of the ``antiquark sea",
 experimentally available in the light front partonic phenomenology. 
In section \ref{sec_observables} we
have shown that the  isospin asymmetry of the sea,  the magnitude of $g_A$ and that of the orbital angular momentum in the nucleon, all indicate that the probability of 5q Fock component 
is  large,  $P_{5q}\sim 0.4$.  One may anticipate even stronger mixing of  baryon resonances, including those which so far are associated with the $qqq$ sector alone. We hope to
address these issues in future publications.

\appendix
\begin{widetext}
\section{Pentaquark variables} \label{sec_5d}
The Jacobi coordinates have been defined in (\ref{eqn_penta_Jacobi}) in terms of four 3d vectors $\vec\alpha, \vec\beta, \vec\gamma, \vec\delta$. For each vector, we use
the usual two polar angles $\theta_i,\phi_i,i=1..4$. The remaining 
variables are the hyperdistance $Y$ and 3 angles in the 4d space, called $\theta2_\chi,\theta1_\chi,\phi_\chi$.
\ba \vec\alpha &=& 
 Y*\rm sin(\theta2\chi)*sin(\theta1\chi)*
   cos(\phi\chi) \times \{sin(\theta\alpha) cos(\phi\alpha), 
    sin(\theta\alpha) sin(\phi\alpha), 
    cos(\theta\alpha) \}  \\
\vec\beta &=& 
 Y*\rm sin(\theta2\chi)*sin(\theta1\chi)*
   sin(\phi\chi) \times \{sin(\theta\beta) cos(\phi\beta), 
    sin(\theta\beta) sin(\phi\beta), cos(\theta\beta) \} \nonumber \\
\vec\gamma &=& 
 Y*\rm sin(\theta2\chi)*
   cos(\theta1\chi) \times \{sin(\theta\gamma) cos(\phi\gamma), 
    sin(\theta\gamma) sin(\phi\gamma), 
    cos(\theta\gamma) \} \nonumber \\ 
\vec\delta &=& 
 Y*\rm cos(\theta2\chi) \times \{sin(\theta\delta) cos(\phi\delta), 
    sin(\theta\delta) sin(\phi\delta), 
    cos(\theta\delta)\} \nonumber \\
\ea
The notations for these 12 variables are
\be y=\{Y, \theta\alpha, \phi\alpha, \theta\beta, \phi\beta, \
\theta\gamma, \phi\gamma, \theta\delta, \phi\delta, \
\theta2\chi, \theta1\chi, \phi\chi \} \ee

The metric tensor is diagonal with its elements (coefficients of $dy_i^2,i=1..12$) given by
\ba
g_{ii}&=&\{1, Y^2 \rm  {cos(\phi\chi)^2} , 
 Y^2 \rm {cos(\phi\chi)^2  sin(\theta1\chi)^2 \
sin(\theta2\chi)^2 sin(\theta\alpha)^2}, \nonumber \\
&& Y^2 \rm  {sin(\theta1\chi)^2 sin(\theta2\chi)^2 \
sin(\phi\chi)^2}, 
 Y^2 \rm{ sin(\theta1\chi)^2 sin(\theta2\chi)^2 sin(\theta\
\beta)^2 sin(\phi\chi)^2},  \nonumber \\
&&
 Y^2 \rm{ cos(\theta1\chi)^2  sin(\theta2\chi)^2},
 Y^2  \rm {cos(\theta1\chi)^2  sin(\theta2\chi)^2 sin(\theta\
\gamma)^2},
 Y^2  \rm {cos(\theta2\chi)^2} , \nonumber \\
&& Y^2  \rm {cos(\theta2\chi)^2  sin(\theta\delta)^2},
 Y^2  ,
 Y^2  \rm {sin(\theta2\chi)^2},
 Y^2   \rm {sin(\theta1\chi)^2 sin(\theta2\chi)^2}\}
\ea
and the volume element 
\be
\sqrt{det(g)}=Y^{11} \rm cos(\theta1\chi)^2 cos(\theta2\chi)^2 \
cos(\phi\chi)^2 sin(\theta1\chi)^4 sin(\theta2\chi)^7 \
sin(\theta\alpha) sin(\theta\beta) sin(\theta\gamma) sin(\
\theta\delta) sin(\phi\chi)^2
\ee

The Laplace-Beltrami operator is 
\ba & -&\vec \nabla^2 = {\partial \over \partial Y^2}  +{11\over Y}{\partial \over \partial Y} +
{1\over Y^2}\rm \big[2 cot(\phi\chi) csc(\theta1\chi)^2 
csc(\theta2\chi)^2 {\partial \over \partial \phi_\chi}
\nonumber \\ &-& 
   \rm 2 csc(\theta1\chi)^2 csc(\theta2\chi)^2 tan(\phi\
\chi) {\partial \over \partial \phi_\chi}
 + 
   csc(\theta1\chi)^2 csc(\theta2\chi)^2 {\partial^2 \over \partial \phi_\chi ^2},
\nonumber \\ &+& 
   \rm 4 cot(\theta1\chi) csc(\theta2\chi)^2 {\partial \over \partial \theta 1_\chi }
 - 
   2 csc(\theta2\chi)^2 tan(\theta1\chi) {\partial \over \partial \theta 1_\chi }
\nonumber \\ &+& ( \rm
 csc(\theta2\chi)^2   {\partial^2 \over \partial \theta 1_\chi^2 }
+ (7 cot(\theta2\chi)   {\partial \over \partial \theta 2_\chi }
- (2 tan(\theta2\chi)   {\partial \over \partial \theta 2_\chi }
+   {\partial^2 \over \partial \theta 2_\chi^2} 
\nonumber \\
&+&  \rm  csc(\theta\delta)^2 sec(\theta2\chi)^2 {\partial^2 \over \partial \phi_\gamma^2}
 + 
   cot(\theta\delta) sec(\theta2\chi)^2 {\partial \over \partial \theta_\gamma}
\nonumber \\
 &+& ( \rm
 sec(\theta2\chi)^2  {\partial^2 \over \partial \theta_\gamma^2}
+   csc(\theta2\chi)^2 csc(\theta\gamma)^2 sec(\theta1_
\chi)^2 {\partial \over \partial \phi_\gamma} 
\nonumber \\
 &+& 
 \rm   cot(\theta\gamma) csc(\theta2\chi)^2 sec(\theta1\
\chi)^2 {\partial \over \partial \theta_\gamma}
 + 
   csc(\theta 2\chi)^2 sec(\theta 1\chi)^2 {\partial^2 \over \partial \theta_\gamma^2}
\nonumber \\
 &+& 
  \rm  csc(\theta1\chi)^2 csc(\theta2\chi)^2 csc(\theta\
\beta)^2 csc(\phi\chi)^2 {\partial^2 \over \partial \phi_\beta^2}
 + 
   cot(\theta\beta) csc(\theta1\chi)^2 \
csc(\theta2\chi)^2 csc(\phi\chi)^2 {\partial \over \partial \theta_\beta} 
\nonumber \\
 &+& 
  \rm  csc(\theta1\chi)^2 csc(\theta2\chi)^2 \
csc(\phi\chi)^2 {\partial^2 \over \partial \theta_\beta^2}
 + 
   csc(\theta1\chi)^2 csc(\theta2\chi)^2 csc(\theta\
\alpha))^2 sec(\phi\chi)^2 {\partial^2 \over \partial \phi_\alpha^2} 
\nonumber 
\\
 &+&   \rm cot(\theta\alpha)) sec(\phi\chi)^2 {\partial \over \partial \theta\alpha}
 +  sec(\phi\chi)^2  {\partial^2 \over \partial \theta\alpha^2} \big]
\ea
is proportional to kinetic energy, provided 
all quarks have the same masses.
\end{widetext}

For spherically symmetric states $\psi=R[Y]$,
only two terms with $Y$ derivative remain. The standard radial wave function redefinition $$R\rightarrow \phi(Y)/Y^{11/2}$$ eliminates the first derivative, but leads to a quasi-centrifugal term
\be -Laplacian= -\phi(Y)^"+{99 \over 4 } {\phi(Y) \over  Y^2}
\ee
For the "L=1" states with the first power of coordinates such as $\vec \alpha$, the (angle averaged) expression becomes 
\be -Laplacian= -\phi(Y)^"+\bigg({99 \over 4 }+10\bigg) {\phi(Y) \over Y^2}
\ee

The interaction between quarks are functions of interquark distances, and thus not spherically symmetric in 12 dimensions. As customary, we evaluate 
them via their angular averaging. As a typical case let us take
the quark pair 1-2, and note that the distance between them is $r_{12}=\sqrt{2}\alpha$. We recall that in the 4d space of lengths $\alpha,\beta,\gamma,\delta$ we have introduced
 three angles $\chi_2,\chi_1,\chi_\phi$, and used them to carry the averaging the of Cornell-like potential as follows
 \be
 \langle r_{12} \rangle={\int \sqrt{2} Y \rm{cos(\chi_2) cos(\chi_2)^2} d\chi_2 \over
 \int \rm {cos(\chi_2)^2 d\chi_2}} ={4 \sqrt{2} \over 3 \pi}Y\approx 0.60 Y
 \ee
 The Coulomb term  diverges logarithmically and needs regularization, so
 the corresponding integration is done till $\chi_2=\pi/2-\epsilon$. We choose 
$\epsilon=0.02$ and obtain
\be \langle{1 \over r_{12}}\rangle\approx {3.3 \over Y}\ee

%%%%%%%%%%%%%%%%%%
\begin{widetext}
\section{Method for finding general pentaquark states} \label{app_algorithmm}

The pentaquark wavefunctions used in this paper were generated using the method described in \cite{Miesch:2024vjk}.  However, a modification must be made to the treatment of pentaquarks described there.  Instead of describing the four quark core and adding on the antiquark, it is possible to jump straight to considering all five particles as long as the \textit{symmetry group} is still $S_4$.  The modified form of the method's "PermuteInBasis" function, now with input $n$ for number of indecies and input $m$ for the order of the symmetry group takes the form

\begin{verbatim}
    (*Basic permutation generators*)
Permutes[n_, m_] :=
 PermutationMatrix[#, n] & /@ GroupGenerators[SymmetricGroup[m]]

 (*Permutes the first m digits in base dim with n total indices*)
ColumnSwap[n_, m_, dim_, i_, x_] := 
 FromDigits[Permutes[n, m][[i]] . IntegerDigits[x, dim, n], dim]
(* needed after flattening all tensors
and i=1 is 12 while i=2 is cycle generator *) 
(*Rank m permuation matrix acting on C^(dim^n)*)
BigPermute[n_, m_, dim_, i_] := 
 SparseArray[
  Table[{x + 1, ColumnSwap[n, m, dim, i, x] + 1} -> 1, {x, 0, 
    dim^n - 1}]]

(*Gives the representation of both permutation generators in the basis\
 spanned by index rearrangements of the tensor T.  \
Now just the first m indices of T*)
PermuteInBasis[T_, m_] := Module[{n, nm, dim, tensorList, basis},
  n = TensorRank[T];
  (*For color nm should be m, 
  for spin it should be n--this has to do with which symmetry group's r\
epresentation the basis is in*)
  If[n - m == 1, nm = n, nm = m];
  dim = Length[T];
  tensorList = 
   Flatten /@ (Transpose[T, PermutationList[#, n]] & /@ 
      GroupElements[SymmetricGroup[nm]]);
  basis = DeleteCases[Orthogonalize[tensorList], {0 ..}];
  {SparseArray[basis . BigPermute[n, m, dim, 1] . Transpose[basis]], 
   SparseArray[basis . BigPermute[n, m, dim, 2] . Transpose[basis]]}
  ]
\end{verbatim}

and the important "GoodBasis" function that gives the basis for the representation of $S_m$ in monoms is

\begin{verbatim}
    (*Gives just the basis associated with T instead*)
GoodBasis[T_, m_] := Module[{n, nm, dim, tensorList, basis},
  n = TensorRank[T];
  (*For spin just goodbasis m doesn't actually matter*)
  If[n - m == 1, nm = n, nm = m];
  tensorList = 
   Flatten /@ (Transpose[T, PermutationList[#, n]] & /@ 
      GroupElements[SymmetricGroup[nm]]);
  SparseArray[DeleteCases[Orthogonalize[tensorList], {0 ..}]]
  ]
\end{verbatim}

The input $T$ to these functions is one Young Tableau with the correct symmetry; for color the natural choice is 

\begin{verbatim}
    pentaColor=TensorProduct[LeviCevita[3],LeviCevita[3]]
\end{verbatim}                                                              

and for each possible total and directional spin/isospin 

\begin{verbatim}
    u = {1, 0};d = {0, 1};
    pentaspin12 = {TensorProduct[d, LeviCivitaTensor[2], 
    LeviCivitaTensor[2]], 
   TensorProduct[u, LeviCivitaTensor[2], LeviCivitaTensor[2]]};
pentaspin32 = {TensorProduct[TensorProduct[d, d, d], 
    LeviCivitaTensor[2]], 
   TensorProduct[Symmetrize[TensorProduct[d, d, u]], 
    LeviCivitaTensor[2]],
   TensorProduct[Symmetrize[TensorProduct[d, u, u]], 
    LeviCivitaTensor[2]],
   TensorProduct[Symmetrize[TensorProduct[u, u, u]], 
    LeviCivitaTensor[2]]};
pentaspin52 = {TensorProduct[d, d, d, d, d],
   Symmetrize[TensorProduct[d, d, d, d, u]],
   Symmetrize[TensorProduct[d, d, d, u, u]],
   Symmetrize[TensorProduct[d, d, u, u, u]],
   Symmetrize[TensorProduct[d, u, u, u, u]],
   TensorProduct[u, u, u, u, u]};
\end{verbatim}

Because the color wavefunction for a pentaquark has \textit{six} indecies (composed of combinations of $\epsilon_{ijk}\epsilon_{lmn}$), for color $n$ is 6 while for spin and isospin it is 5.  In both cases however, $m$ is still 4. This doesn't cause any issues, because the matrices found with PermuteInBasis all belong to representations the same group $S_m$ and therefore can always be Kronecker producted together into a new representation.

The coordinate parts of the wavefunctions can be treated essentially the same as \cite{Miesch:2024vjk}, except the permutations matrices acting on them are now $n$-dimensional matrices of $m$-dimensional permutations:

\begin{verbatim}
    (*Transforms to modified jacobi coordinates*)
Jacobi[n_] := 
 Inverse[DiagonalMatrix[
    Table[Sqrt[i + 1]/Sqrt[i], {i, 1, n}]]] . (Table[
     Join[Table[1/i, i], Table[0, n - i]], {i, 1, n}] + 
    DiagonalMatrix[Table[-1, n - 1], 1])
(*Permutation matricies in jacobi basis*)
JacobiPermute[n_, m_, i_] := 
 Transpose@
  SparseArray[(Jacobi[n] . Permutes[n, m][[i]] . Inverse[Jacobi[n]])[[
    1 ;; n - 1, 1 ;; n - 1]]]
\end{verbatim}

Clearly an explicit example is in order here. In Table  \ref{tab_L0} we have seen that the simplest states are for maximal spin $S=5/2$ (to be discussed in the next subsection)
while the {\em most complex} case is for $S=I=1/2$, with the largest dimension of the ``good basis" 
$3\times 5\times 5=75$ for color,spin and isospin, respectively. These dimensions can readily be understood in terms of Young tableaux, for example 3 for color corresponds to three of them shown in (\ref{Y1}).

Two generators of the permutation group $S_4$ are: (1) $G_1$ is the interchange of particles $1\leftrightarrow 2$; (2) $G_2$ is a cycling one, moving all indices by one.
For the most complex example of $L=0$ pentaquarks at hand,
with  $S=1/2,I=1/2,S_z=-1/2$, these two permutation generators  take the form
$$ G_1= \begin{tabular}{|ccc|}
   -1  &  0 & 0\\
  0  & -1 & 0\\
  0 & 0 & 1\\
  \end{tabular}
  \otimes 
  \begin{tabular}{|ccccc|}
   1/2  &  0 & $\sqrt{3}/2$ & 0 & 0\\
  0  & 1/2 & 0 & $\sqrt{3}/2$ & 0 \\
  $\sqrt{3}/2$ & 0 & -1/2 & 0 & 0\\
  0 & $\sqrt{3}/2$ & 0 & -1/2 & 0 \\
  0& 0& 0& 0& 1\\ \end{tabular}
  \otimes 
  \begin{tabular}{|ccccc|}
   1/2  &  0 & $\sqrt{3}/2$ & 0 & 0\\
  0  & 1/2 & 0 & $\sqrt{3}/2$ & 0 \\
  $\sqrt{3}/2$ & 0 & -1/2 & 0 & 0\\
  0 & $\sqrt{3}/2$ & 0 & -1/2 & 0 \\
  0& 0& 0& 0& 1\\ \end{tabular}
 $$

$$ G_2=\begin{tabular}{|ccc|} 
1/3 & 2$\sqrt{2}/3$ & 0 \\
-$\sqrt{2}/3$ & 1/6 & $\sqrt{3}/2$ \\
$\sqrt{2/3}$ & -1/2$\sqrt{3}$ & 1/2\\
 \end{tabular}
  \otimes 
   \begin{tabular}{|ccccc|}
   -1/4 & -$\sqrt{3}/4$ & $\sqrt{3}/4$ & 3/4 & 0\\
$\sqrt{3}/4$ & -1/4 & 1/4 & -1/4$\sqrt{3}$ & $\sqrt{2}/3$ \\
-$\sqrt{3}/4$ & -3/4 &-1/4 & -$\sqrt{3}/4$ & 0 \\
3/4 & -$\sqrt{3}/4$ & -1/4$\sqrt{3}$ & 1/12 & -$\sqrt{2}/3$ \\
   0 & 0 & $\sqrt{2/3}$ & -$\sqrt{2}/3$ & -1/3\\
   \end{tabular}
    \otimes 
   \begin{tabular}{|ccccc|}
      -1/4 & -$\sqrt{3}/4$ & $\sqrt{3}/4$ & 3/4 & 0\\
$\sqrt{3}/4$ & -1/4 & 1/4 & $-1/4\sqrt{3}$ & $\sqrt{2}/3$ \\
-$\sqrt{3}/4$ & -3/4 &-1/4 & -$\sqrt{3}/4$ & 0 \\
3/4 & -$\sqrt{3}/4$ & $-1/4\sqrt{3}$ & 1/12 & $-\sqrt{2}/3$ \\
   0 & 0 & $\sqrt{2/3}$ & $-\sqrt{2}/3$ & -1/3\\
   \end{tabular}
$$
One can then proceed to do the tensor product, after which simultaneously diagonalize these two generators (e.g. with the AntiSymmetricStates method), and 
find 3 common antisymmetric vectors for both 75-dimensional matrices. Those are three pentaquark states. Returning from
``good basis" notations back to large monom space, one have 
their wave functions explicitly.
\end{widetext}

%%%%%%%%%%%%%%%%%%
\section{ The pentaquark states built via explicit Young tableaux} \label{app_analytic}
For readers who found the previous subsection to be too quick,
let us show how one builds the states using Young tableaux explicitly.
The space-color-spin-flavor representations of
the pentaquark states $[qqqq\bar q]$ with light $u,d$ flavors are involved, owing to the anti-symmetrization  of the core $[qqqq]$. In this Appendix, we briefly detail their construction using representations of the
permutation group of $S_4$ in addition to the
standard unitary representations. The construction makes extensive use of the character representations of $S_4$, to extract the pertinent representations. It is a natural extension of the construction for the select set of $S_3$ representations  introduced by Isgur and Karl~\cite{Isgur:1978xj,Isgur:1978wd}, and extended later to pentaquarks in~\cite{An:2005cj,an:2006zf,Xu:2014yya,Duan:2016rkr,An:2019tld}.
The results are in overall agreement with the monom construction detailed above. \\

\subsection{Color and flavor representations}
The color singlet wavefunction of $q^4\bar q$ follows from the combination of the color triplet of $q^4$ with the color anti-triplet of $\bar q$. Since the total wavefunction is antisymmetric in  color,  the corresponding Young tableau follows from the product 
\begin{widetext}
\bea
{\bf 1}_C=\bigg(
\begin{ytableau}
$$&\\
&\\
&\\
\end{ytableau}\equiv C[222]\bigg)=
\bigg(
\begin{ytableau}
$$&\cr \cr \cr
\end{ytableau}\equiv C[211]\bigg)\otimes
\bigg(
\begin{ytableau}
$$\cr \cr
\end{ytableau}\equiv C[11]\bigg)
\eea
\end{widetext}
with the quark-antiquark color identifications for the Young tableaux $C[222]=[q^4\bar q]_C$,
$C[211]=[q^4]_C$ and $C[11]=[\bar q]_C$.
The color representation $C[211]$ is 3-degenerate mixed representation
\bea
\label{Y1}
%[q^4]_C=
\begin{ytableau}
1&2\\3\\4\\
\end{ytableau}
\qquad
\begin{ytableau}
1&3\\2\\4\\
\end{ytableau}
\qquad
\begin{ytableau}
1&4\\2\\3\\
\end{ytableau}
\eea

To garentee the anti-symmetry of the full color-space-spin-flavor wavefunction, the space-spin-flavor wavefunction $[q^4]_{[LFS]}$ for the identical quarks is valued in the conjugate representation of its color counterpart $[q^4]_C$, 
\bea
[q^4]_{LSF}=\overline{[q^4]_C}=
\begin{ytableau}
$$& &\\
\\
\end{ytableau}\equiv LSF[31]
\eea
is 3-degenerate as well
\bea
%[q^4]_{LSF}=
\begin{ytableau}
1&3 & 4\\2\\
\end{ytableau}
\qquad
\begin{ytableau}
1&2 & 4\\3\\
\end{ytableau}
\qquad
\begin{ytableau}
1&2 & 3\\4\\
\end{ytableau}
\eea

\subsection{Jacobi-like representations of $S_4$}
Before proceeding to the representations of $S_4$ we first recall the Jacobi coordinates for 5 particles, which will be used to generalize the Isgur-Karl 
representations of the 3 particles wavefunctions to 5 particles. Another form of   Jacobi coordinates we use is
\bea
\vec\alpha&=&\frac 1{\sqrt 2} \vec r_{12}\nonumber\\
\vec\beta&=&\frac 1{\sqrt 6} (\vec r_{13}+\vec r_{23})\nonumber\\
\vec\gamma&=&\frac 1{\sqrt 12} (\vec r_{14}+\vec r_{24}+\vec r_{34})\nonumber\\
\vec\delta&=&\frac 1{\sqrt 20} (\vec r_{15}+\vec r_{25}+\vec r_{35}+\vec r_{45})
\eea
with $\vec r_{1,2,3,4}$referring to the coordinates of the 4 quarks,
and $\vec r_5$ to the coordinate of the fifth anti-quark. The CM coordinate is given by
\bea
\vec R=\frac 15 (\vec r_1+\vec r_2+\vec r_3+\vec r_4+\vec r_5)
\eea

\subsection{Color representations}
The first 2 Young tableaux in (\ref{Y1}) are readily identified with the two irreducible Jacobi-like $\rho=\alpha, \lambda=\beta$ representations of $S_3$ 
which are parts of the cycles in  $S_4$, and which we have previously used~\cite{Miesch:2024vjk}. More specifically,
\bea
\label{Y2}
C[211]_\alpha=
\begin{ytableau}
1&3\\2\\4\\
\end{ytableau}
%\equiv |C[211](3121)\rangle_\alpha
\eea
\bea
C[211]_\beta  =
\begin{ytableau}
1&2\\3\\4\\
\end{ytableau}
%\equiv |C[211](3211)\rangle_\beta
\eea
\bea
C[211]_\gamma=
\begin{ytableau}
1&4\\2\\3\\
\end{ytableau}\nonumber\\
%\equiv |C[211](1321)\rangle_\gamma
\eea
%with the rightmost notation referring to the  standard basis for the Young Tableaux.
Note that the removal
of box 4 in the first two tableaux, yields the mixed symmetry representation for 3 identical quarks $[q^3]_M$  which is 2-degenerate, as  per the Isgur-Karl representation~\cite{Miesch:2024vjk}.
The last Young tableau  is commensurate with the Jacobi-like $\gamma$, as the new irreducible representation of $S_4$.

To make more explicit the color representations $[211]_{\xi=\alpha, \beta, \gamma}$, we
use the standard  $C=R,G,B$ and $\bar C=\bar R,\bar G,\bar B$ for the quark and anti-quark colors respectively.
From the Young tableau color representation, it follows that the singlet pentaquark state
follows from pairing the valence anti-quark color with the detached Yoing box, to neutralize 
the color. More explicitly, we have
\begin{widetext}
\bea
\label{SINGLET211}
C[211]_\xi=\frac 1{\sqrt 3}\big(
C[221]_\xi(R)\bar R+ C[221]_\xi(G)\bar G+ C[221]_\xi(B)\bar B\big)
\eea
with 
\bea
\label{COLOR211X}
C[211]_\alpha(C)=
\begin{ytableau}
1&3\\2\\4\\
\end{ytableau}\,\,
\begin{ytableau}
R&C\\G\\B\\
\end{ytableau}
\qquad
C[211]_\beta(C)=
\begin{ytableau}
1&2\\3\\4\\
\end{ytableau}\,\,
\begin{ytableau}
R&C\\G\\B\\
\end{ytableau}
\qquad
C[211]_\gamma(C)=
\begin{ytableau}
1&4\\3\\2\\
\end{ytableau}\,\,
\begin{ytableau}
R&C\\G\\B\\
\end{ytableau}\qquad
\eea
The expicit forms of the Jacobi color states, are obtained by using the pertinent projection operators of the permutation group of $S_4$, on each of the  $[211](R,G,B)$ representations~\cite{Yan:2012zzb}. With this in mind, we have explicitly for $C=R$
\bea
\label{COLOR211R}
C[211]_\alpha(R)=
\frac 1{\sqrt 48}&&
\big(3|RGRB\rangle-3|GRRB\rangle-3|RBRG\rangle+3|BRRG\rangle
-|RGBR\rangle \nonumber\\
&&+|GRBR\rangle +|RBGR\rangle -|BRGR\rangle
+2|GBRR\rangle-2|BGRR\rangle\big)\nonumber\\
%%%%
C[211]_\beta(R)=
\frac 1{\sqrt 16}&&
\big(2|RRGB\rangle-2|RRBG\rangle-|RGRB\rangle-|GRRB\rangle
+|RBRG\rangle \nonumber\\
&&+|BRRG\rangle +|RGBR\rangle +|GRBR\rangle
-|RBGR\rangle-|BRGR\rangle\big)\nonumber\\
%%%%
C[211]_\gamma(R)=
\frac 1{\sqrt 6}&&
\big(|BRGR\rangle+|RGBR\rangle+|GBRR\rangle-|RBGR\rangle
-|GRBR\rangle -|BGRR\rangle\big)
\eea
The other color representations with $C=G,B$ in (\ref{SINGLET211}),
are obtained similarly, with the results 
\bea
\label{COLOR211G}
C[211]_\alpha(G)=
\frac 1{\sqrt 48}&&
\big(3|RGGB\rangle-3|GRGB\rangle-3|BGGR\rangle+3|GBGR\rangle
-|RGBG\rangle \nonumber\\
&&+|GRBG\rangle -|GBRG\rangle +|BGRG\rangle
+2|BRGG\rangle-2|RBGG\rangle\big)\nonumber\\
%%%%
C[211]_\beta(G)=
\frac 1{\sqrt 16}&&
\big(2|GGBR\rangle-2|GGRB\rangle+|RGGB\rangle+|GRGB\rangle
-|RGBG\rangle \nonumber\\
&&-|GRBG\rangle +|GBRG\rangle +|BGRG\rangle
-|GBGR\rangle-|BGGR\rangle\big)\nonumber\\
%%%%
C[211]_\gamma(G)=
\frac 1{\sqrt 6}&&
\big(|RGBG\rangle-|GRBG\rangle-|RBGG\rangle+|BRGG\rangle
+|GBRG\rangle -|BGRG\rangle\big)
\eea
and
\bea
\label{COLOR211B}
C[211]_\alpha(B)=
\frac 1{\sqrt 48}&&
\big(3|BRBG\rangle-3|RBBG\rangle+3|GBBR\rangle-3|BGBR\rangle
+|RBGB\rangle \nonumber\\
&&-|BRGB\rangle -|GBRB\rangle +|BGRB\rangle
+2|RGBB\rangle-2|GRBB\rangle\big)\nonumber\\
%%%%
C[211]_\beta(B)=
\frac 1{\sqrt 16}&&
\big(2|BBRG\rangle-2|BBGR\rangle+|RBGB\rangle+|BRGB\rangle
-|GBRB\rangle \nonumber\\
&&-|BGRB\rangle -|RBBG\rangle -|BRBG\rangle
+|GBBR\rangle+|BGBR\rangle\big)\nonumber\\
%%%%
C[211]_\gamma(B)=
\frac 1{\sqrt 6}&&
\big(|RGBB\rangle-|GRBB\rangle-|RBGB\rangle+|BRGB\rangle
+|GBRB\rangle -|BGRB\rangle\big)
\eea
It is readily checked that the colored core states (\ref{COLOR211R}-\ref{COLOR211B}) are ortho-normalized for each color sector $R,G,B$. 
We note that (\ref{COLOR211R})  is in  agreement with~\cite{Yan:2012zzb,Xu:2014yya}.

%\end{widetext}
\subsection{Spin, flavor: $[q^4]_{S,F}$}
The spin flavor configurations of $[q^4]$ follow from
the standard $SU(2)$ Young tableaux, 
%\begin{widetext}
\bea
\label{PRODUCT1}
[q^4]_S=
\begin{ytableau}
$$&&&\\
\end{ytableau}
\,\oplus\,
\begin{ytableau}
$$& &\\ \\
\end{ytableau}\,\oplus\,
\begin{ytableau}
$$& \\ &\\
\end{ytableau}\equiv S[4]\oplus  S[31]\oplus  S[22]
\eea
and similarly for $[q^4]_F$ with $S\rightarrow F$,

\subsubsection{Maximum weight representations for $[q^4]_{S,F}$}
We now construct explicitly the 4 quark states with maximum spin 
that appear in (\ref{PRODUCT1}) by using the standard procedures 
for $SU(2)_S$ Young tableaux. For the completly symmetric tableau,
the state is unique
\bea
S[4]^{22}=
\begin{ytableau}
\uparrow & \uparrow & \uparrow  & \uparrow\\
\end{ytableau}=
\uparrow \uparrow \uparrow \uparrow 
\eea
For the mixed tableau,
\bea
S[31]^{11}=
\begin{ytableau}
\uparrow & \uparrow & \uparrow\\
\downarrow\\
\end{ytableau}\rightarrow S[31]_{\alpha,\beta,\gamma}^{11}
\eea
there are 3 properly symmetrized states. These states are readily obtained
from the Jacobi-like $\rho,\lambda$ Isgur-Karl spin states discussed in~\cite{Miesch:2024vjk}, by
tagging an additional spin state. In the $\alpha,\beta,\gamma$ labeling used here, they are
\bea
S[31]_\beta^{11}&=&\frac 1{\sqrt 2}(\uparrow\downarrow-\downarrow\uparrow)\uparrow\uparrow\nonumber\\
S[31]_\alpha^{11}&=&\frac 1{\sqrt 6}(
2\uparrow\uparrow\downarrow
-\uparrow\downarrow\uparrow-\downarrow\uparrow\uparrow)\uparrow
\nonumber\\
S[31]_\gamma^{11}&=&\frac 1{\sqrt{12}}
(3\uparrow\uparrow\uparrow\downarrow
-\downarrow\uparrow\uparrow\uparrow
-\uparrow\downarrow\uparrow\uparrow
-\uparrow\uparrow\downarrow\uparrow)
\nonumber\\
\eea
The explicit construction of $S[22]^{00}$ follows a similar reasoning,
\bea
S[22]^{00}=
\begin{ytableau}
\uparrow & \uparrow \\
\downarrow & \downarrow\\
\end{ytableau}\rightarrow S[22]^{00}_{\alpha, \beta}
\eea
with  2 symmetrized states.
Note that the maximum representations for the core $[q^4]_F$ follow a similar reasoning  for $SU(2)_F$ with $u,d$ light flavors.

\subsubsection{Maximum weight representations for $([q^4]\otimes\bar q)_{S,F}$}
The  maximum weight configuration for the spin in pentastates are obtained by recoupling through pertinent Clebsch-Gordon coefficients, 
\bea
\label{S545252}
S_5[4]\bigg[\frac 52\frac 52\bigg]=|\uparrow\uparrow\uparrow\uparrow\bar\uparrow\rangle
\eea
%%%
and
%%%%%%
\bea
S_5[4]\bigg[\frac 32\frac 32\bigg]=
\sqrt{\frac 45}
|\uparrow\uparrow\uparrow\uparrow\bar\downarrow\rangle
-\sqrt{\frac 1{20}}\big(
|\uparrow\uparrow\uparrow\downarrow\bar\uparrow\rangle
+|\uparrow\uparrow\downarrow\uparrow\bar\uparrow\rangle
+|\uparrow\downarrow\uparrow\uparrow\bar\uparrow\rangle
+|\downarrow\uparrow\uparrow\uparrow\bar\uparrow\rangle
\big)
\eea
%%%%%
and
%%%
\bea
S_5[22]_\alpha\bigg[\frac 12\frac 12\bigg]&=&
\sqrt{\frac 14}\big(|\uparrow\downarrow\uparrow\downarrow\bar\uparrow\rangle
-|\downarrow\uparrow\uparrow\downarrow\bar\uparrow\rangle
-|\uparrow\downarrow\downarrow\uparrow\bar\uparrow\rangle
+|\downarrow\uparrow\downarrow\uparrow\bar\uparrow\rangle\big)
\nonumber\\
S_5[22]_\beta\bigg[\frac 12\frac 12\bigg]&=&
\sqrt{\frac 1{12}}\big(
2|\uparrow\uparrow\downarrow\downarrow\bar\uparrow\rangle
-|\uparrow\downarrow\uparrow\downarrow\bar\uparrow\rangle
-|\downarrow\uparrow\uparrow\downarrow\bar\uparrow\rangle
-|\uparrow\downarrow\downarrow\uparrow\bar\uparrow\rangle
-|\downarrow\uparrow\downarrow\uparrow\bar\uparrow\rangle
+2|\downarrow\downarrow\uparrow\uparrow\bar\uparrow\rangle\big)
\eea
%%%%%
and
%%%
\bea
\label{S5313232}
S_5[31]_\alpha\bigg[\frac 32\frac 32\bigg]&=&\sqrt{\frac 12}
\big(|\uparrow\downarrow\uparrow\uparrow\bar\uparrow\rangle
-|\downarrow\uparrow\uparrow\uparrow\bar\uparrow\rangle
\big)
\nonumber\\
S_5[31]_\beta\bigg[\frac 32\frac 32\bigg]&=&\sqrt{\frac 16}
\big(2|\uparrow\uparrow\downarrow\uparrow\bar\uparrow\rangle
-|\uparrow\downarrow\uparrow\uparrow\bar\uparrow\rangle
-|\downarrow\uparrow\uparrow\uparrow\bar\uparrow\rangle
\big)
\nonumber\\
S_5[31]_\gamma\bigg[\frac 32\frac 32\bigg]&=&\sqrt{\frac 1{12}}
\big(3|\uparrow\uparrow\uparrow\downarrow\bar\uparrow\rangle
-|\uparrow\uparrow\downarrow\uparrow\bar\uparrow\rangle
-|\uparrow\downarrow\uparrow\uparrow\bar\uparrow\rangle
-|\downarrow\uparrow\uparrow\uparrow\bar\uparrow\rangle
\big)
\eea
%%%%%
and
%%%%%%
\bea
\label{S5311212}
S_5[31]_\alpha\bigg[\frac 12\frac 12\bigg]&=&
\sqrt{\frac 13}
\big(|\uparrow\downarrow\uparrow\uparrow\bar\downarrow\rangle
-|\downarrow\uparrow\uparrow\uparrow\bar\downarrow\rangle\big)
-\sqrt{\frac 1{12}}
\big(|\uparrow\downarrow\uparrow\downarrow\bar\uparrow\rangle
+|\uparrow\downarrow\downarrow\uparrow\bar\uparrow\rangle
-|\downarrow\uparrow\uparrow\downarrow\bar\uparrow\rangle
-|\downarrow\uparrow\downarrow\uparrow\bar\uparrow\rangle\big)
\nonumber\\
S_5[31]_\beta\bigg[\frac 12\frac 12\bigg]&=&
{\frac 13}
\big(2|\uparrow\uparrow\downarrow\uparrow\bar\downarrow\rangle
-|\uparrow\downarrow\uparrow\uparrow\bar\downarrow\rangle
-|\downarrow\uparrow\uparrow\uparrow\bar\downarrow\rangle
\big)\nonumber\\
&&-{\frac 1{6}}
\big(2|\uparrow\uparrow\downarrow\downarrow\bar\uparrow\rangle
-|\uparrow\downarrow\uparrow\downarrow\bar\uparrow\rangle
-|\downarrow\uparrow\uparrow\downarrow\bar\uparrow\rangle
+|\uparrow\downarrow\downarrow\uparrow\bar\uparrow\rangle
+|\downarrow\uparrow\downarrow\uparrow\bar\uparrow\rangle
-2|\downarrow\downarrow\uparrow\uparrow\bar\uparrow\rangle
\big)
\nonumber\\
S_5[31]_\gamma\bigg[\frac 12\frac 12\bigg]&=&
\sqrt{\frac 1{18}}
\big(3|\uparrow\downarrow\uparrow\uparrow\bar\downarrow\rangle
-|\downarrow\uparrow\uparrow\uparrow\bar\downarrow\rangle
-|\uparrow\uparrow\downarrow\uparrow\bar\downarrow\rangle
-|\uparrow\uparrow\uparrow\downarrow\bar\downarrow\rangle
\big)\nonumber\\
&&-\sqrt{\frac 1{18}}
\big(|\uparrow\uparrow\downarrow\downarrow\bar\uparrow\rangle
+|\uparrow\downarrow\uparrow\downarrow\bar\uparrow\rangle
+|\downarrow\uparrow\uparrow\downarrow\bar\uparrow\rangle
-|\uparrow\downarrow\downarrow\uparrow\bar\uparrow\rangle
-|\downarrow\uparrow\downarrow\uparrow\bar\uparrow\rangle
-|\downarrow\downarrow\uparrow\uparrow\bar\uparrow\rangle
\big)
\eea
Note that the anti-quark spin satisfies
$\bar\uparrow =-i\sigma^2\uparrow=\uparrow$ 
and $\bar\downarrow =-i\sigma^2\downarrow=-\downarrow$.
Finally, we note that the 
maximum weight representations for the flavor contributions 
follow from a similar reasoning, with the substitutions
$\uparrow\rightarrow u\,,\downarrow\rightarrow d\,.$

\subsection{S-state pentaquarks}
To proceed to the $LSF$ representations, we will first focus on the ground pentaquark or S-states with $L=0$, and then proceed to detail the extension of the construction to the P-states which is more involved.

\subsubsection{Spin-flavor mixed representations: $SF[31]$}
For the ground state wavefunction with $L=0$, the 4-quarks spin-flavor mixed representation conjugate to the 4-quarks color representation is also 3-degenerate, with the Jacobi-like nomenclature
\bea
SF[31]_\alpha=
\begin{ytableau}
1&3 & 4\\2\\
\end{ytableau}
%\equiv|SF[31](1121)\rangle_\alpha
\eea
\bea
SF[31]_\beta=
\begin{ytableau}
1&2 & 4\\3\\
\end{ytableau}
%\equiv|SF[31](1211)\rangle_\beta
\eea
\bea
SF[31]_\gamma=
\begin{ytableau}
1&2 & 3\\4\\
\end{ytableau}
%\equiv|SF[31](2111)\rangle_\gamma
\eea

These mixed representations follow from the product representations 
\bea
\label{PRODUCT}
[q^4]_S\otimes [q^4]_F
\eea
with a mixed $[31]_{S,F}$ net component. It follows 
from the character representations of $S_4$ that the product representations (\ref{PRODUCT}) using (\ref{PRODUCT1}) are 
\bea
S[4]\otimes F[31]&=&SF[31]\nonumber\\
S[31]\otimes F[4]&=&SF[31]\nonumber\\
S[31]\otimes F[31]&=&SF[4]\oplus SF[31]\oplus SF[22]\nonumber\\
S[31]\otimes F[22]&=&SF[31]\nonumber\\
S[22]\otimes F[22]&=&SF[4]\oplus SF[22]
\eea
 and the spin-flavor exchanged products
\bea
F[4]\otimes S[31]&=&SF[31]\nonumber\\
F[31]\otimes S[31]&=&SF[4]\oplus SF[31]\oplus SF[22]\nonumber\\
F[31]\otimes S[22]&=&SF[31]\nonumber\\
F[31]\otimes S[4]&=&SF[31]
\eea
hence
\bea
SF[31]\subset  S[4]\otimes F[31], S[31]\otimes F[4], S[31]\otimes F[31], S[31]\otimes F[22], S[22]\otimes F[31]
\eea

The properly anti-symmetrized $[31]_{FS}$ Jacobi core states are
\bea
\label{XX0}
(SF[31]:[31]_F\otimes [4]_S )_{\alpha, \beta, \gamma}&=&
F[31]_{\alpha, \beta, \gamma}S[4]\nonumber\\
(SF[31]:[4]_F\otimes [31]_S )_{\alpha, \beta, \gamma}&=&
F[4]S[31]_{\alpha, \beta, \gamma}
\eea
and
\bea
\label{XX1}
(SF[31]:[31]_F\otimes [31]_S )_\alpha&=&
\frac 1{\sqrt 3}\bigg(F[31]_\alpha S[31]_\beta+\alpha\leftrightarrow \beta\bigg)+\frac 1{\sqrt 6}\bigg(F[31]_\alpha S[31]_\gamma+\alpha\leftrightarrow \gamma\bigg)\nonumber\\
(SF[31]:[31]_F\otimes [31]_S )_\beta&=&
\frac 1{\sqrt 3}\bigg(F[31]_\alpha S[31]_\alpha-\alpha\rightarrow \beta\bigg)+\frac 1{\sqrt 6}\bigg(F[31]_\gamma S[31]_\beta+\gamma\leftrightarrow \beta\bigg)\nonumber\\
(SF[31]:[31]_F\otimes [31]_S )_\gamma&=&
\frac 1{\sqrt 6}\bigg(F[31]_\alpha S[31]_\alpha+\alpha\rightarrow \beta\bigg)-\frac 2{\sqrt 6}F[31]_\gamma S[31]_\gamma
\eea
and
\bea
\label{XX2}
(SF[31]:[22]_F\otimes [31]_S )_\alpha&=&
\frac 1{2}\bigg(F[22]_\alpha S[31]_\beta+\alpha\leftrightarrow \beta\bigg)-\frac 1{\sqrt 2}F[22]_\alpha S[31]_\gamma\nonumber\\
(SF[31]:[22]_F\otimes [31]_S )_\beta&=&
\frac 1{2}\bigg(F[22]_\alpha S[31]_\alpha-\alpha\rightarrow \beta\bigg)-\frac 1{\sqrt 2}F[22]_\beta S[31]_\gamma\nonumber\\
(SF[31]:[22]_F\otimes [31]_S )_\gamma&=&
-\frac 1{\sqrt 2}\bigg(F[22]_\alpha S[31]_\alpha+\alpha\rightarrow \beta\bigg)
\eea
and
\bea
\label{XX3}
(SF[31]:[31]_F\otimes [22]_S )_\alpha&=&
\frac 1{2}\bigg(S[22]_\alpha F[31]_\beta+\alpha\leftrightarrow \beta\bigg)-\frac 1{\sqrt 2}S[22]_\alpha F[31]_\gamma\nonumber\\
(SF[31]:[31]_F\otimes [22]_S )_\beta&=&
\frac 1{2}\bigg(S[22]_\alpha F[31]_\alpha-\alpha\rightarrow \beta\bigg)-\frac 1{\sqrt 2}S[22]_\beta F[31]_\gamma\nonumber\\
(SF[31]:[31]_F\otimes [22]_S )_\gamma&=&
-\frac 1{\sqrt 2}\bigg(S[22]_\alpha F[31]_\alpha+\alpha\rightarrow \beta\bigg)
\eea
%\end{widetext}

\subsubsection{ Pentaquarks, S-shell ($L=0$) and maximal spin $S=5/2$}
In this work we consider the  case of pentaquarks formed of only the light $u,d$  flavors of equal mass. The ground state wavefunction carries $L=0$ or space symmetric S-state.
As a result, the totally antisymmetric color-spin-flavor wavefunction in the S-state reads
%\begin{widetext}
\bea
\label{PENTA1}
\Psi^S_P&=&\frac {\varphi_{00_L}}{\sqrt 3}\,
\bigg(\bigg[
C[211]_\beta SF[31]_\alpha-C[211]_\alpha SF[31]_\beta+C[211]_\gamma\,SF[31]_\gamma\bigg]C[11] SF[1]\bigg)
\eea
\end{widetext}
with $\varphi_{00_L}(R)$ valued in terms of the hyper-spherical distance $R$.
Note that the reduction of (\ref{PENTA1}) to the 3 particle subspace
is of the  anti-symmetric form $$C_\beta SF_\alpha-C_\alpha SF_\beta$$
familiar from the singlet ${\bf 1}_A$ Isgur-Karl representation for the nucleon odd-parity excited state (see Eq. A4 in~\cite{Miesch:2024vjk}).  (\ref{PENTA1}) is in agreement with the result in~\cite{Xu:2014yya}.

Following the strategy used previously for 6-q hexoquark states, we start
with the maximal spin case, in which the spin part of the wave function is trivially symmetric. Now it means that spin is $S=5/2$. The remaining
wave functions are the $SU(3)$ color and $SU(2)$ flavor (isospin), and
one may think that following the usual representation theory based on Young
tableaux it can be done analytically, this is indeed the case, but due to  its extended nature we put derivation in Appendix ??, see (\ref{}).

Since the state
with maximal spin $and$ isospin $S=I=5/2$ cannot produce antisymmetric
state consistent with Fermi statistics, we have two such states possible,
with $I=3/2$ or $I=1/2$.

%\subsubsection{ Pentaquarks, S-shell general}

In Table~\ref{tab_LX} we list the $L=0$ pentaquark states, with the 
degeneracy associated to the pertinet spin-isospin entry, in agreement with the monom construction above.

\begin{table}[h!]
    \centering
    \begin{tabular}{|c|c|c|c|} \hline
  I/S    & 1/2 & 3/2   & 5/2  \\  \hline
  1/2 & 3  & 3    & 1 \\
  3/2 & 3  & 3  &  1   \\
  5/2 & 1  & 1  & 0  \\ \hline
    \end{tabular}
    \caption{Pentaquark states
    with $L=0$, spin $S$ and isospin $I$.  }
    \label{tab_LX}
\end{table}

%\subsubsection{Explicit example}
For a direct comparison to the monom construction above, we explicit here the
S-pentas states with isospin $\frac 12$ and spin $\frac 52$ content
\begin{widetext}
\bea
&&\varphi_P^S\Bigg[ \frac 12\frac 12,\frac 52\frac 52\Bigg]=
\frac{\varphi_{00L}}{\sqrt 3}|\uparrow \uparrow \uparrow \uparrow \bar\uparrow \rangle\nonumber\\
&&\times\bigg(
C[211]_\beta \frac 1{\sqrt 2}(ud-du)uu\bar u  -
C[211]_\alpha \frac 1{\sqrt 6}(2uud-udu-duu)u\bar u \nonumber\\
&&\qquad +
C[211]_\gamma \frac 1{\sqrt 12}(3uuud-duuu-uduu-uudu)\bar u\bigg)
\eea
with the color entries $C[211]_{\xi=\alpha,\beta,\gamma}$ given in (\ref{SINGLET211}).
After averaging over the spatial part, one readily checks that the
individual isospin contributions $i=1,2,3,4$ in the core are equal as they should
\bea
\varphi_P^S\Bigg[ \frac 12\frac 12,\frac 52\frac 52\Bigg]^\dagger I_i^3
\varphi_P^S\Bigg[ \frac 12\frac 12,\frac 52\frac 52\Bigg]=\frac 14
\eea
Similarly, the total 3-spin, total 3-isospin, and  total isospin  are
\bea
\varphi_P^S\Bigg[ \frac 12\frac 12,\frac 52\frac 52\Bigg]^\dagger \bigg(\sum_{i=1}^5S_i^3\bigg)
\varphi_P^S\Bigg[ \frac 12\frac 12,\frac 52\frac 52\Bigg]=\frac 52\nonumber\\
\varphi_P^S\Bigg[ \frac 12\frac 12,\frac 52\frac 52\Bigg]^\dagger \bigg(\sum_{i=1}^5I_i^3\bigg)
\varphi_P^S\Bigg[ \frac 12\frac 12,\frac 52\frac 52\Bigg]=\frac 12\nonumber\\
\varphi_P^S\Bigg[ \frac 12\frac 12,\frac 52\frac 52\Bigg]^\dagger \bigg(\sum_{i=1}^5(\vec I_i)^2\bigg)
\varphi_P^S\Bigg[ \frac 12\frac 12,\frac 52\frac 52\Bigg]=\frac 34
\eea
 the 12-pair color and isospin correlations are
\bea
\varphi_P^S\Bigg[ \frac 12\frac 12,\frac 52\frac 52\Bigg]^\dagger \bigg(\lambda_1\cdot\lambda_2\bigg)
\varphi_P^S\Bigg[ \frac 12\frac 12,\frac 52\frac 52\Bigg]&=&
\varphi_P^S\Bigg[ \frac 12\frac 12,\frac 52\frac 52\Bigg]^\dagger \frac 13\bigg(\vec C_4^2-\frac {16}3\bigg)
\varphi_P^S\Bigg[ \frac 12\frac 12,\frac 52\frac 52\Bigg]
=-\frac 43\nonumber\\
%%%
%%%
\varphi_P^S\Bigg[ \frac 12\frac 12,\frac 52\frac 52\Bigg]^\dagger \bigg(\tau_1\cdot\tau_2\bigg)
\varphi_P^S\Bigg[ \frac 12\frac 12,\frac 52\frac 52\Bigg]&=&
\varphi_P^S\Bigg[ \frac 12\frac 12,\frac 52\frac 52\Bigg]^\dagger 
\bigg(\frac 13\vec I_4^2-1\bigg)
\varphi_P^S\Bigg[ \frac 12\frac 12,\frac 52\frac 52\Bigg]
=-\frac 13
\eea
%%%
%%%
and the spin-spin correlations are
\bea
\varphi_P^S\Bigg[ \frac 12\frac 12,\frac 52\frac 52\Bigg]^\dagger \bigg(\sum_{i<j}S_i\cdot S_j\bigg)
\varphi_P^S\Bigg[ \frac 12\frac 12,\frac 52\frac 52\Bigg]&=&
\varphi_P^S\Bigg[ \frac 12\frac 12,\frac 52\frac 52\Bigg]^\dagger 
\bigg(\frac 12\vec S^2-\frac{15}8\bigg)
\varphi_P^S\Bigg[ \frac 12\frac 12,\frac 52\frac 52\Bigg]
=\frac 52
\eea

\subsubsection{Pentaquarks in P-shell $L=0$ and $FS[31]$ core}
%\subsectionautorefnamesection{ P-state pentaquarks }
The pentaquark P-states commensurate with the nucleon state $\frac 12\frac 12^+$, are more numerous and involved in their $LSF$ arrangements. 
The P-state pentaquark with an S-state symmetric core and an 
antiquark in a P-state, is 
%\begin{widetext}
\bea
\label{PENTA2}
\Psi^A_{Pm_L}=\frac{\varphi_{m_L}}{\sqrt 3}
\bigg(\bigg[
C[211]_\beta ([4]_L[31]_{SF})_\alpha
-C[211]_\alpha ([4]_L[31]_{SF})_\beta
+C[211]_\gamma ([4]_L[31]_{SF})_\gamma\bigg]C[11] SF[1]\bigg)
\nonumber\\
\eea
\end{widetext}
with the short hand notation
\bea
([4]_L[31]_{SF})_{\alpha,\beta,\gamma}=
[4]_L([31]_{SF})_{\alpha,\beta,\gamma}
\eea
Note that the mixed Jacobi $SF$ representations follow from
3 distinct arrangements
\bea
&&[4]_L([31]_{FS}:[31]_F\otimes [31]_S)\nonumber\\
&&[4]_L([31]_{FS}:[22]_F\otimes[31]_S)\nonumber\\
&&[4]_L([31]_{FS}:[31]_F\otimes[22]_S)
\eea
making the state (\ref{PENTA2}) for fixed $m_L$, 3-degenerate.
The ensuing and explicit Jacobi states are given in (\ref{XX1}-\ref{XX3}).

\subsubsection{P-states:  $L=1$ and $FS[4]$ core }
%\subsubsection{$L\otimes SF$}
From parity considerations, it follows that for the core $[q^4]$ only $L[4]$
with the unpaired $\bar q$ in a P-state, or
$L[31]$ with the unpaired $\bar q$ in an S-state are allowed. Using the character assignments for finite groups, and the tethrahedral  representations of $S_4$, it follows that for 2 flavors, the mixed  representations are
\bea
L[4]\otimes SF[31]&=&LSF[31]\nonumber\\
L[31]\otimes SF[31]&=&LSF[4]\oplus LSF[31]\oplus LSF[22]\nonumber\\
L[31]\otimes SF[22]&=&LSF[31]\nonumber\\
L[31]\otimes SF[4]&=&LSF[31]
\eea

The pentaquark with a spatial P-core but with a  'symmetric' FS-core, to which we can tag an anti-quark in an  S-state, is given by
\begin{widetext}
\bea
\label{PENTA3}
\Psi^P_P&=&\frac {\tilde\varphi_{1m_L}}{\sqrt 3}\,
\bigg(\bigg[
C[211]_\beta ([31]_L[4]_{FS})_\alpha-C[211]_\alpha ([31]_L[4]_{FS})_\beta+C[211]_\gamma([31]_L[4]_{FS})_\gamma\bigg]C[11] SF[1]\bigg)\nonumber\\
\eea
\end{widetext}
with $\tilde \varphi_{1m_L}(R)$ valued in terms of the hyper-spherical distance $R$. The 'symmetric' FS-core is composed of  2 mixed Jacobi configurations
\bea
&&([31]_L([4]_{FS}:[31]_F[31]_S))_{\alpha,\beta,\gamma}
\nonumber\\
&&([31]_L([4]_{FS}:[22]_F[22]_S))_{\alpha,\beta,\gamma}
\eea

\subsubsection{P-states:  $L=1$ and $FS[31]$ core }
There are 3 mixed Jacobi  configurations with a spatial P-core but with an  'asymmetric' FS-core 
\bea
&&([31]_L([31]_{FS}:[31]_F[31]_S))_{\alpha,\beta,\gamma}\nonumber\\
&&([31]_L([31]_{FS}:[22]_F[31]_S))_{\alpha,\beta,\gamma}\nonumber\\
&&([31]_L([31]_{FS}:[31]_F[22]_S))_{\alpha,\beta,\gamma}
\eea

\subsection{Spin-color-flavor operators and Hamiltonians}
In the S-state the spin-orbit and tensor interactions vanish, with only the
spin-color exchange from the perturbative gluon exchange, and the emergent 't Hooft flavor induced interactions adding to the central kinetic and confining interactions. We now consider them sequentially.

\subsubsection{Color interaction}
The simplest pair color interaction in the pentaquark state reads
\bea
\label{ZEROTH}
&&\langle \Psi^S_P|\sum_{i<j}\lambda_i\cdot \lambda_j|\Psi^S_P\rangle=\nonumber\\
&&\langle \Psi^S_P|\bigg(2\vec C_5^2-\frac {40}3\bigg)|\Psi^S_P\rangle=-\frac {40}3
\eea
where the Casimir operator $\vec C_5^2=\sum_{i=1}^5\lambda_i^2/4$  is  null in the color singlet pentaquark state. Note that when the exchange
involves the antiquark labeled by 5,  the replacement of the Gelman matrices
$\lambda_5\rightarrow -\lambda^*$ is assumed (and similarly $\sigma_5\rightarrow -\sigma^*$ for the spin below).

\subsubsection{Perturbative color-spin interaction}
We use the same protocol when evaluating the color-spin interaction induced by 1-gluon exchange in
the pentaquark S-state, for light quarks of mass $m_Q$. The pair interactions 
\bea
\label{PERT}
&&\mathbb V_{1g}(1,2,3,4,5)=\nonumber\\
&&\sum_{i<j\leq 5}V_{1g}(r_{ij})\frac {\lambda_i\cdot\lambda_j\,\sigma_i\cdot\sigma_j}{m_Q^2}
\eea
are sandwitched between the
pertinent color-spin wavefunctions in (\ref{PENTA1}).

Assuming the short ranged $V_{1g}$ to be about constant, we can simplify the color-spin interaction by noting that the 4 quark core $[q^4]$ is antisymmetric under the 
 spin-flavor-color permutations
\bea
P^S_{ij}P^F_{ij}P^C_{ij}[q^4]=-[q^4]
\eea
 with $i,j=1,2,3,4$ and 
\bea
P^S_{ij}&=&\frac 12 (1+\sigma_i\cdot\sigma_j)\nonumber\\
P^F_{ij}&=&\frac 12 (1+\tau_i\cdot\tau_j)\nonumber\\
P^C_{ij}&=&\frac 12 (\frac 23+\lambda_i\cdot\lambda_j)
\eea
hence the identity
\bea
\label{CASIMIR1}
-\sum_{i<j}^4\lambda_i\cdot\lambda_j
\sigma_i\cdot \sigma_j=
-\frac {32}3+\frac 43 \vec{S}^2_4+4\vec{I}^2_4+2\vec{C}^2_4\nonumber\\
\eea
Here 
\bea
\vec S^2_4&=&\sum_{i=1}^4 \frac 14\vec \sigma_i^2\nonumber\\
\vec I^2_4&=&\sum_{i=1}^4 \frac 14\vec \tau_i^2\nonumber\\
\vec C^2_4&=&\sum_{i=1}^4 \frac 14\vec \lambda_i^2
\eea
are respectively, the spin, isospin and color
Casimirs for the  completly antisymmetric 4-quark  configuration $[q^4]$.
We further note that the color Casimir for the antisymmetric core is that of the fundamental color triplet representation with $\vec C_4^2\rightarrow \frac 43$, hence
\bea
\label{CASIMIR1X}
-\sum_{i<j}^4\lambda_i\cdot\lambda_j
\sigma_i\cdot \sigma_j\rightarrow
-8+\frac 43 \vec{S}^2_4+4\vec{I}^2_4
\eea

\subsubsection{Color-spin shift from  the core}
The pentaquark states are configurated in a core plus valence antiquark
$[q^4\bar q]$. The expectation value of (\ref{CASIMIR1X})  depends solely on the the net spin and isospin of the core irrespective of the $\xi=\alpha, \beta, \gamma$ assignments from 
permutation symmetry, hence for (\ref{XX1}-\ref{XX3}) we have respectively
\begin{widetext}
\bea
\label{FIRST}
\langle \Psi^S_P|\mathbb V_{1g}|\Psi^S_P\rangle_4=-\frac {V_{1g}}{3m_Q^2}
\sum_{\xi=\alpha, \beta, \gamma} 
\big(SF_5[31]_\xi\bigg(-8+\frac 43 \vec{S}^2_4+4\vec{I}^2_4\bigg)
SF_5[31]_\xi\big)
%=\frac {4V_{1g}}{3m_Q^2}(11, 17,13)
\eea
\end{widetext}

\subsubsection{Color-spin shift from the valence}
The remaining spin-color exchange operator between the pairs $i5$ (i-quark  and 5-antiquark) with $i=1,2,3,4$ in the core, can be obtained by  evaluating
\bea
\label{SS45}
\lambda_4\cdot \lambda_5
\,\sigma_4\cdot\sigma_5 =2\bigg(\vec C_{4+5}^2-\vec C_4^2-\vec C_5^2\bigg)
\,\sigma_4\cdot\sigma_5\nonumber\\
\eea
in terms of the pertinent Casimirs for the $4-5$ particles, 
and then using the permutation symmetry of the core to deduce
the other combinations through the identity
\begin{widetext}
\bea
\lambda_i\cdot \lambda_5
\,\sigma_i\cdot\sigma_5=
P^S_{i4}P^C_{i4}\bigg[\lambda_4\cdot \lambda_5
\,\sigma_4\cdot\sigma_5\bigg]P^C_{i4}P^S_{i4}\rightarrow
P^F_{i4}\bigg[\lambda_4\cdot \lambda_5
\,\sigma_4\cdot\sigma_5\bigg]P^F_{i4}=
\lambda_4\cdot \lambda_5
\,\sigma_4\cdot\sigma_5
\eea
for $i=1,2,3$. 
The rightmost result holds only for expectation values in Pentastates with an antisymmetric core, and follows from the 
commutation of the flavor matrices with the spin-color matrices.

The shift induced by (\ref{SS45})
in the nucleon-like pentaquark states ,
splits into  
\bea
\label{SECOND}
\langle \Psi^S_P|\mathbb V_{1g}|\Psi^S_P\rangle_5=\frac {4V_{1g}}{3m_Q^2}
\bigg[
&&+\big(C_5[211]_\alpha \lambda_4\cdot\lambda_5 C_5[211]_\alpha\big)\times
\big(SF_5[31]_\beta  \sigma_4\cdot\sigma_5 SF_5[31]_\beta \big)
\nonumber\\
&&+\big(C_5[211]_\beta\lambda_4\cdot\lambda_5 C_5[211]_\beta\big)\times
\big(SF_5[31]_\alpha  \sigma_4\cdot\sigma_5 SF_5[31]_\alpha\big)
\nonumber\\
&&+\big(C_5[211]_\gamma \lambda_4\cdot\lambda_5 C_5[211]_\gamma\big)\times
\big(SF_5[31]_\gamma  \sigma_4\cdot\sigma_5 SF_5[31]_\gamma \big)
\big)
%+ 4\rightarrow 1,2,3
\qquad
\bigg]
%\nonumber\\
\eea
where only the diagonal contributions are non-zero due
to the diagonal form of the color matrix element. Indeed, 
the color contributions in (\ref{SECOND}) are readily obtained by noting that the 4-5-particles are either in an octet in  the $\alpha, \beta$ color configuration, or a singlet in the $\gamma$ color representation, hence
\bea
\label{C5211}
C_5[211]_{\alpha,\beta} \lambda_4\cdot\lambda_5 C_5[211]_{\alpha,\beta}&=&
2\big(C_{8_c}^2-C_{3_c}^2-C_{\bar 3_c}^2\big)
=2\bigg(3-\frac 43-\frac{4}3\bigg)=\frac 23
\nonumber\\
C_5[211]_{\gamma} \lambda_4\cdot\lambda_5 C_5[211]_{\gamma}
&=&
2\big(C_{1_c}^2-C_{3_c}^2-C_{\bar 3_c}^2\big)=
2\bigg(0-\frac 43-\frac {4}3\bigg)=-\frac{16}3
\eea
where $C^2_{X_c}$ are the color Casimirs for the pertinent color representation $X_c$ of $SU(3)_c$. Recall that for the $SU(3)_c$  color representation $X_c$ with Dynkin index $D(p,q)$ (with p-quarks and q-antiquarks), the quadratic Casimir is 
\bea
C^2_{X_c}=\frac 13 (p^2+q^2+3p+3q+pq)
\eea
The recoupling of the 4-5-spins  in (\ref{SECOND}) is more involved, and require the explicit spin configurations for the pentaquark states. For the maximum weight, the Jacobi spin
states are given in (\ref{S545252}-\ref{S5311212}).
\\
\\
\noindent {\bf $\bf L=0, I=\frac 12, S=\frac 52$ pentastate:}
\\
For $L=0$ there is  1 pentastate with $IJ=\frac 12\frac 52$.
The 45-spin contribution (\ref{SECOND}) 
for $\xi=\alpha, \beta, \gamma$ amounts to
\bea
\label{SF531X}
SF_5[31]_\xi  \sigma_4\cdot\sigma_5 SF_5[31]_\xi=
\big(S[4]\sigma_4\cdot\sigma_5 S[4]\big)\,
\big(F[31]_\xi F[31]_\xi\big)=1
\eea
and similarly for $4\rightarrow 1,2,3$ thanks to the symmetry of the spin combination $S[4]$. The combination of (\ref{C5211})  and (\ref{SF531X}) in (\ref{SECOND}), yields
\bea
\label{SECONDZ}
\langle \Psi^S_P\bigg[\frac 12\frac 52 \bigg]|\mathbb V_{1g}|\Psi^S_P\bigg[\frac 12\frac 52\bigg]\rangle_5=
\frac{4V_{1g}}{3m_Q^2}\bigg(2\times \frac 23-\frac{16}3=-4\bigg)
\eea
The final result for the perturbative gluon exchange (\ref{PERT}) in this  pentastate, is the sum of the 4-core contribution
\bea
\label{FIRST}
\langle \Psi^S_P\bigg[\frac 12\frac 52\bigg]|\mathbb V_{1g}|\Psi^S_P\bigg[\frac 12\frac 52\bigg]\rangle_4=-\frac {V_{1g}}{3m_Q^2}
\sum_{\xi=\alpha, \beta, \gamma} 
\big(S[4]F[31]_\xi\bigg(-8+\frac 43 \vec{S}^2_4+4\vec{I}^2_4\bigg)
S[4]F[31]_\xi\big)_4=\frac {V_{1g}}{m_Q^2}\bigg(-8\bigg)\nonumber\\
\eea
plus the 5-remainder valence contribution (\ref{SECONDZ}), 
\bea
\label{FINALXX1252}
\langle \Psi^S_P\bigg[\frac 12\frac 52\bigg]|\mathbb V_{1g}|\Psi^S_P\bigg[\frac 12\frac 52\bigg]\rangle_{4+5}=
\frac {4V_{1g}}{3m_Q^2}
\bigg(-4-6=-10\bigg)\nonumber\\
\eea
in agreement with the result derived using the monon basis analysis listed in the main text.
\\
\\
\noindent {\bf $\bf L=0, I=\frac 12, S=\frac 32$ pentastate:}
\\
For $L=0$, there are 3 pentastates with $IJ=\frac 12\frac 32$ assignments with the following cores as given in  (\ref{XX1}-\ref{XX3})
\bea
\label{CORE1232}
C[211](S[31]F[31]), C[211](S[31]F[22]), C[211](S[4]F[31])
\eea
The color-spin contributions from the cores are respectively
\bea
\label{FIRST}
\langle \Psi^S_P|\mathbb V_{1g}|\Psi^S_P\rangle_4=-\frac {V_{1g}}{3m_Q^2}
\sum_{\xi=\alpha, \beta, \gamma} 
\big(SF_5[31]_\xi\bigg(-8+\frac 43 \vec{S}^2_4+4\vec{I}^2_4\bigg)
SF_5[31]_\xi\big)=\frac {V_{1g}}{m_Q^2}\bigg(-\frac 83, +\frac{16}3,-8\bigg)
\eea
Once combined with the valence contributions, the results are in agreement with the monon 
basis analysis listed in the main text.
%The nucleon with three light flavors $[q^3]$ could have a pentaquark admixture
%$[q^4\bar q]$ in its ground state with net spin-isospin assignment $SI=\frac 12 \frac 12$. 
%There are 3 distinct nucleon-like pentaquark states with these assignments 
%A rerun of the preceding arguments, yields
%\bea
%\label{FINALXX1212}
%\langle \Psi^S_P\bigg[\frac 12\frac 32\bigg]|\mathbb V_{1g}|\Psi^S_P\bigg[\frac 12\frac 32\bigg]\rangle_{4+5}=
%\frac {4V_{1g}}{3m_Q^2}
%\bigg(-2+1=-1, 4-5=-1, -6+6=0\bigg)
%\eea
%with the core plus the remainder shown for each of the cores in (\ref{CORE1232}) sequentially.
\\
\\
\noindent {\bf $\bf L=0, I=\frac 12, S=\frac 12$ pentastate:}
\\
For $L=0$, there are 3 pentastates with $IJ=\frac 12\frac 12$ assignments with the following cores
\bea
\label{CORE1212}
C[211](S[31]F[31]), C[211](S[31]F[22]), C[211](S[22]F[31])
\eea
%\begin{widetext}
The color-spin contributions from the cores are respectively
\bea
\label{FIRST}
\langle \Psi^S_P|\mathbb V_{1g}|\Psi^S_P\rangle_4=-\frac {V_{1g}}{3m_Q^2}
\sum_{\xi=\alpha, \beta, \gamma} 
\big(SF_5[31]_\xi\bigg(-8+\frac 43 \vec{S}^2_4+4\vec{I}^2_4\bigg)
SF_5[31]_\xi\big)=\frac {V_{1g}}{m_Q^2}\bigg(-\frac 83, +\frac{16}3,0\bigg)
\eea
Again, once combined with the valence contributions, the results are in agreement with the monon 
basis analysis listed  in the main text.
%\end{widetext}
%Combining the valence contribution (\ref{C5211}) with the core contribution (\ref{FIRST}), yields
%\bea
%\label{FINALXX1212}
%\langle \Psi^S_P\bigg[\frac 12\frac 12\bigg]|\mathbb V_{1g}|\Psi^S_P\bigg[\frac 12\frac 12\bigg]\rangle_{4+5}=
%\frac {4V_{1g}}{3m_Q^2}
%\bigg(-2+1=-1, 4-12=-8, 0+0=0\bigg)
%\eea
%with the core plus the remainder shown for each of the cores in (\ref{CORE1212}) sequentially.
%\end{widetext}

\subsubsection{Non-perturbative 't Hooft interactions}
The non-perturbative  't Hooft interactions induced by rescattering of light quark zero modes through instantons and anti-instantons is
\bea
\label{THOOFT}
&&\mathbb V_{TH}(1,2,3,4,5)=\nonumber\\
&&\sum_{i,j\leq 5}V_{TH}(r_{ij})
(1-\tau_i\cdot\tau_j)(1-a\sigma_i\cdot\sigma_j)
\eea
with
\bea
%V_{TH}(r_{ij})\rightarrow 
V_{TH}=-\frac 14|\kappa_2|A_{2N}=-|\kappa_2|\frac{(2N_c-1)}{2N_c(N_c^2-1)}\nonumber\\
\eea
for zero size instantons, and $a=\frac 1{2N_c-1}\rightarrow \frac 15$ which is seen to vanish  in the large $N_c$ limit. In this limit, (\ref{THOOFT}) can be readily evaluated 
in the nucleon-like pentaquark states 
%\begin{widetext}
\bea
\langle \Psi^S_P|\mathbb V_{TH}|\Psi^S_P\rangle=
\frac {V_{TH}}3\sum_{\xi=\alpha, \beta, \gamma}
 SF_5[31]_\xi \bigg(\sum_{i,j\leq 5}
(1-\tau_i\cdot\tau_j)
%(1-a\sigma_i\cdot\sigma_j)
\bigg)
SF_5[31]_\xi= V_{TH}(26,26,26)
\eea
\end{widetext}

{\bf Acknowledgements}. 
This work is supported by the Office of Science, U.S. Department of Energy under Contract  No. DE-FG-88ER40388.
It is also supported in part within the framework of the Quark-Gluon Tomography (QGT) Topical Collaboration, under contract no. DE-SC0023646.

\bibliography{penta2}
\end{document}